\documentstyle[12pt,amssymb,amsfonts,amsbsy,html]{article}
\begin{document}
\author{Ilja Schmelzer\thanks
       {WIAS Berlin}}

 \title{General Ether Theory}
\sloppypar 

\begin{abstract}
The paper is an introduction into General Ether Theory
(GET).  We start with few assumptions about an universal ``ether'' in
a Newtonian space-time which fulfils

\begin{eqnarray*}
\partial_t \rho + \partial_i (\rho v^i) &= &0 \\
\partial_t (\rho v^j) + \partial_i(\rho v^i v^j + p^{ij}) &= &0
\end{eqnarray*}

For an ``effective metric'' $g_{\mu\nu}$ we derive a Lagrangian where
the Einstein equivalence principle is fulfilled:

\[
L = L_{GR}
  - (8\pi G)^{-1}(\Upsilon g^{00}-\Xi (g^{11}+g^{22}+g^{33}))\sqrt{-g}
\]

We consider predictions (stable frozen stars instead of black holes,
big bounce instead of big bang singularity, a dark matter term),
quantization (regularization by an atomic ether, superposition of
gravitational fields), related methodological questions (covariance,
EPR criterion, Bohmian mechanics).

\end{abstract}

\maketitle
\tableofcontents
\newtheorem{theorem}{Theorem}
\newtheorem{axiom}{Axiom}

\section{Introduction}

The purpose of the present work is to present an alternative metric
theory of gravity.  The Lagrangian of the theory

\[
L = L_{GR}
  - (8\pi G)^{-1}(\Upsilon g^{00}-\Xi (g^{11}+g^{22}+g^{33}))\sqrt{-g}
\]

is very close to the GR Lagrangian, and in the limit $\Xi,\Upsilon\to
0$ we obtain the classical Einstein equations.

The key point is that this Lagrangian may be {\em derived \/} starting
with a few assumptions about the ``ether'' -- a classical medium in a
classical Newtonian background with Euclidean space and absolute time
${\Bbb R}^3\otimes{\Bbb R}$.  We need only a few general principles: a
Lagrange formalism and its relation with standard conservation laws.
The gravitational field $g^{\mu\nu}$ is defined by the ``general''
steps of freedom of the ether -- density $\rho$, velocity $v^i$,
pressure $p^{ij}$.\footnote{As usual, we use latin indices for
three-dimensional indices and greek indices for four-dimensional
indices.  We also use the notation $\hat{g}^{\mu\nu} =
{g}^{\mu\nu}\sqrt{-g}$.}  The matter fields describe its material
properties.  What explains the Einstein equivalence principle is that
the ether is universal: all fields describe properties of the ether,
there is no external matter.  Therefore, observers are also only
excitations of the ether, unable to observe some of the ether
properties.  This explains that we are unable to observe all steps of
freedom of the ether.  We need no artificial conspiracy or highly
sophisticated model to obtain relativistic symmetry in an ether
theory.

The only difference to GR are two additional terms which depend on the
preferred coordinates $X^i, T$.  In ``weak'' covariant formulation
(with the preferred coordinates handled as ``fields'' $X^i(x), T(x)$)
we obtain:

\[L = L_{GR}-(8\pi G)^{-1}(
   	         \Upsilon g^{\mu\nu}T_{,\mu}T_{,\nu}
       	       - \Xi      g^{\mu\nu}\delta_{ij}X^i_{,\mu}X^j_{,\nu}
   	)\sqrt{-g}
\]

Instead of no equation for the preferred coordinates, we obtain a
well-defined general equation for these coordinates: the classical
conservation laws, which appear to be the harmonic coordinate
condition.

But this changes a lot.  It is, essentially, a paradigm shift as
described by Kuhn \cite{Kuhn}.  We revive the metaphysics of Lorentz
ether theory in its full beauty.  This requires the reconsideration of
the whole progress made in fundamental physics in this century.
Therefore, after derivation of the theory and their comparison with
existing theories of gravity we reconsider different domains of
science from point of view of the new paradigm.  Of course, this may
be only a raw overview, a program for future research instead of a
summary of results.  But this raw overview does not suggest serious
problems for the new paradigm, while essential problems of the
relativistic paradigm disappear.

The preferred background leads to well-defined local energy and
momentum conservation laws.  Moreover, the additional terms seem to be
useful to solve cosmological problems. The $\Xi$-term defines a nice
homogeneous dark matter candidate.  The $\Upsilon$-term is even more
interesting: it avoids the big bang singularity and leads instead to a
bounce.  Such a bounce makes the cosmological horizon much larger and
therefore solves the cosmological horizon problem without inflation.
This term stops also the gravitational collapse immediately before
horizon formation.  Because of the underlying Euclidean symmetry, the
flat universe is the only homogeneous universe.  Therefore, GET is not
only in agreement with observation, but allows to solve some
serious cosmological problems solved today by inflation theory.

A very strong argument in favour of GET is the violation of Bell's
inequality.  In our opinion, it is a very simple and decisive proof of
the existence of a preferred foliation -- as simple and decisive as
possible in fundamental physics.  Only if we try to avoid this simple
conclusion, the issue becomes complicate -- we have to reject simple
fundamental principles like the EPR criterion of reality or causality.
In our opinion, there is a lot of confusion in this question.  For
example, it is often assumed that the EPR criterion is in
contradiction with quantum theory.  But the existence of Bohmian
mechanics proves that there is no such contradiction.

In quantum field theory the reintroduction of a preferred frame does
not lead to problems. Instead, it allows to generalize Bohmian
mechanics into the relativistic domain and clarifies the choice of the
Fock space in semiclassical field theory.

The ``ether hypothesis'' suggests also a simple solution for the
problem of non-renormalizability.  Technically, this solution is
already known as ``effective field theory''.  In this concept, it is
assumed that below a certain cutoff scale the theory becomes really
different.  This concept has two features where GET suggests
modification: First, in the standard concept the nature of the theory
below this cutoff remains completely unspecified.  Instead, GET
suggests a well-defined framework: canonical quantum theory, Newtonian
space-time, and some ``atomic ether theory''.  Second, the cutoff
length is supposed to be the Planck length.  Instead, GET makes a
prediction which is inconsistent with Planck length -- the cutoff is
defined by $g^{00}\sqrt{-g}V_{cutoff}=1$. As a consequence, the cutoff
length seems to increase in a homogeneous ``expanding'' universe.

Remarkably, all these results are only side-effects.  It was not the
original intention of the author to revive old ether theory.  Instead,
the author shares the common admiration for the beauty of GR.  It was
also not the intention to solve cosmological problems -- they have
been considered only after the derivation of the Lagrangian.  We also
have not tried to save the EPR criterion or to generalize Bohmian
mechanics.

The original motivation was different.  It was a quantum gravity
thought experiment which has convinced the author that a Newtonian
framework is necessary.  The question is if a ``one world theory'' as
GR is sufficient to describe superpositions of different gravitational
fields, or if such a superposition depends on relations between the
superposed fields, their ``relative position''.  To decide this
question, we consider a simple interaction of a superposition of
quasiclassical fields with a test particle.  We observe a transition
probability which depends on such relative information.  To describe
such transition probabilities in quantum gravity it seems necessary to
introduce a common background.

Last not least, it seems necessary to criticize some aspects of the
relativistic paradigm for quantum gravity.  ``Because of the lack of
data, quantum gravity is strongly influenced by philosophical
prejudices of the researchers'' \cite{Butterfield}, therefore, these
prejudices have to be considered.  We use Rovelli \cite{Rovelli} as a
base for this consideration.  It includes an excellent methodological
part we agree with.  We criticize his relativistic argumentation and
argue that our consideration is in much better agreement with the
proposed methodology.

\section{General Ether Theory}

Let's now define general ether theory.  We have a Newtonian framework
-- absolute Euclidean space with orthonormal coordinates $X^i$ and
absolute time T.  We have also classical causality -- causal influence
$A\to B$ between events A and B is possible only if $T(A)\le T(B)$.

The ether is described by steps of freedom which are usual in
condensed matter theory: there is an ``ether density'' $\rho(X,T)$, an
``ether velocity'' $v^i(X,T)$ and an ``ether pressure'' $p^{ij}(X,T)$.
As usual for a density, $\rho>0$.

These steps of freedom define the gravitational field.  The theory is
a metric theory of gravity, and the metric $g_{\mu\nu}$ is defined
algebraically by the following formula:

\begin{eqnarray*} \label{gdef}
g^{00} \sqrt{-g} &=& \rho \\
g^{i0} \sqrt{-g} &=& \rho v^i \\
g^{ij} \sqrt{-g} &=& \rho v^i v^j + p^{ij}
\end{eqnarray*}

This formula is a variant of the ADM decomposition.  Especially,
$v^i$ is the ADM shift vector.

Because the density $\rho$ is always positive, this formula defines a
Lorentz metric if and only if the tensor $p^{ij}$ is negative
definite.  Therefore, we make the additional assumption that $p^{ij}$
is negative definite.

\subsection{The material properties of the ether}

These are not all steps of freedom of the ether.  Instead, there are
other steps of freedom, the ``material properties'' $\varphi^m(X,T)$
of the ether.  But these steps of freedom are not defined by GET.
Instead, GET is a general theory only, it describes only a few general
properties, not all properties of the ether.  It is, therefore, a
meta-theory, a general scheme for an ether theory.  Different ether
theories can fit into this scheme.  Ether theories with well-defined
material properties and material laws we name ``complete ether
models''.

This is in no way strange for a theory of gravity.  All metric
theories of gravity are general schemes in the same sense.  They do
not specify the matter steps of freedom and the matter Lagrangian.
Nonetheless, they specify an essential and very important property of
the matter Lagrangian -- that it fulfils the Einstein equivalence
principle.  Thus, the meta-theoretical character is a common feature
of theories of gravity.

GET in some sense explains this subdivision into a universal
gravitational field and matter fields.  Indeed, there is a similar
subdivision in condensed matter theory -- the subdivision between the
few basic steps of freedom (like density, velocity, pressure) which
are common for very different materials and the ``material
properties'' which differ for different materials.  In GET this
subdivision fits with the subdivision into gravity and matter fields:
Density, velocity and pressure are used to describe the gravitational
field, while the other material properties describe the matter fields.

This is an essential difference to classical ether theory.  In the
classical concept, the ether is assumed to be something different from
usual matter.  In GET, usual matter is described by continuous fields,
and these fields describe various properties of the ether.

But we not only assume that there are material properties of the ether
described by some matter fields.  We assume more: all matter fields
describe material properties of the ether.  There is no ethr-external
matter:

\begin{axiom}[universality] \label{universality_axiom}
There is nothing except the ether.  All fields describe steps of
freedom of the ether.
\end{axiom}

Thus, the complete ether model is the theory of everything.

\subsection{Conservation laws}

In our covariant formalism, the conservation laws are the
Euler-Lagrange equations for the preferred coordinates $X^\mu$ (see
(\ref{covariant_conservation})).  Now, let's try to find these
conservation laws.  The main hypothesis is that these conservation
laws coinside with the classical conservation laws we know from
condensed matter theory.

\begin{axiom}[continuity equation] \label{continuity_axiom}
The mass of the ether is conserved.  This conservation is
described by the classical continuity equation:
\begin{equation} \label{continuity}
 \partial_t \rho + \partial_i (\rho v^i) = 0
\end{equation}
\end{axiom}

The other important equation of classical condensed matter theory is
the Euler equation.  It is the conservation law for momentum:

\begin{axiom}[Euler equation] \label{Euler_axiom}
The momentum of the ether is conserved.  This conservation is
described by the classical Euler equation:

\begin{equation} \label{Euler}
 \partial_t (\rho v^j) + \partial_i(\rho v^i v^j + p^{ij})=0
\end{equation}
\end{axiom}

Note that we have no terms for external forces or interaction with
external matter.  The reason is that we have already incorporated here
the universality axiom -- there is no momentum exchange with external
matter, there are no external forces. All ``matter fields'' are
``material properties'' of the ether.

These are already all ether equations specified by GET.  All other
equations are ``material laws'' of the ether, they depend on the
``material properties'' $\varphi^m$ which have to be defined only
by the complete ether model.  GET does not specify them.

Now, a key observation is what happens if if we rewrite the classical
conservation laws as equations for the effective metric $g_{\mu\nu}$.
We obtain a well-known equation -- the harmonic condition:

\[ \Box X^\nu = \partial_\mu (g^{\mu\nu}\sqrt{-g}) = 0\]

\subsection{Lagrange formalism}

A major assumption is that we have a Lagrange formalism.  We use the
covariant formulation of the theory: the preferred coordinates $X^\mu$
are considered as fields, the Lagrangian depends on the fields
$X^\mu(x)$ in a covariant way (see \S{}\ref{covariant_formalism}).  In
this formalism, the conservation laws are the Euler-Lagrange equations
for the preferred coordinates (see (\ref{covariant_conservation})).

On the other hand, we have already found the conservation laws and
observed that they may be written as equations for the preferred
coordinates.  Thus, it seems reasonable to assume that they are
proportional:

\[ {\delta S\over\delta X^\mu} = \gamma_\mu \Box X^\mu\]

Now, the coefficients $\gamma_\mu$ may be different.  We can use
Euclidean symmetry to argue that $\gamma_1=\gamma_2=\gamma_3$, but
there is no reason to suppose a relation between $\gamma_0$ and the
$\gamma_i$.

Instead of the $\gamma_\mu$ we introduce introduce a diagonal matrix
$\gamma_{\mu\nu}$ with $\gamma_{\mu\mu} = -4\pi G \gamma_\mu$.  The
factor $4\pi G$ is well-known from GR, and it seems natural to
introduce it here: in this case, the two constants
$\Upsilon=\gamma_{00}$, $\Xi=-\gamma_{ii}$ appear in a similar way as
Einstein's cosmological constant $\Lambda$.  Now we can formulate the

\begin{axiom}[Lagrange formalism] \label{Lagrange_axiom}
There exists a ``weak covariant'' Lagrange formalism so that the
Euler-Lagrange equations for the preferred coordinates $X^\mu$ and the
classical conservation laws for the ether $\Box X^\mu=0$ are related
in the following way:

\begin{equation} \label{proportionality_condition}
{\delta S\over\delta X^\mu} = -(4\pi G)^{-1}\gamma_{\mu\nu}\Box X^\nu
\end{equation}
\end{axiom}

Now, let's find the general Lagrangian which fulfils this assumption.

\begin{theorem}
The general Lagrangian for GET is

\begin{equation} \label{GET-Lagrangian}
L = -(8\pi G)^{-1} \gamma_{\mu\nu}g^{\mu\nu} \sqrt{-g}
  + L_{GR}(g_{\mu\nu}) + L_{matter}(g_{\mu\nu},\varphi^m)
\end{equation}
\end{theorem}

Proof: First, we find a Lagrangian which fulfils the condition
(\ref{proportionality_condition}):

\[L_{GET} = -(8\pi G)^{-1}\gamma_{\mu\nu}X^\mu_{,\alpha}X^\nu_{,\beta}
  	g^{\alpha\beta}\sqrt{-g}
\]

For the difference we obtain

\[ {\delta \int(L-L_{GET})\over\delta X^\mu} = 0 \]

Thus, the remaining part is ``strong'' covariant, that means, it is
not only covariant, but does not depend on the preferred coordinates
$X^\mu$ too.  But this is the classical requirement for the Lagrangian
of general relativity.  Thus, we can identify the difference with the
classical Lagrangian of general relativity.

\[
L = L_{GET}(g_{\mu\nu},X^\mu)
  + L_{GR}(g_{\mu\nu}) + L_{matter}(g_{\mu\nu},\varphi^m)
\]

In the preferred coordinates $L_{GET}$ may be rewritten as

\[L_{GET}=-(8\pi G)^{-1}(\Upsilon g^{00}-\Xi(g^{11}+g^{22}+g^{33}))\sqrt{-g}\]

or in a more compact form as

\[L_{GET}=-(8\pi G)^{-1} \gamma_{\mu\nu}g^{\mu\nu} \sqrt{-g} \]

This proves the theorem.

It should be noted that we have no theoretical reason to fix the signs
for the cosmological constants $\Xi,\Upsilon,\Lambda$.  Their values
should be fixed by observation.

\section{Simple properties}

Now we have defined GET and can describe its properties.

First, let's write down the other Euler-Lagrange equations.  As
equations for $g^{\mu\nu}$ we obtain the Einstein equations with two
additional non-covariant terms:

\begin{equation}\label{GET-Einstein}
G^\mu_\nu  = 8\pi G (T_m)^\mu_\nu
   + (\Lambda +\gamma_{\kappa\lambda}g^{\kappa\lambda}) \delta^\mu_\nu
   - 2g^{\mu\kappa}\gamma_{\kappa\nu}
\end{equation}

As in GR, the equations for the matter fields $\varphi^m$ depend on
the matter Lagrangian and remain unspecified.

The expression of the GET Lagrangian in terms of the original ether
variables is quite nice:

\[ 4\pi G \Xi^{-1} L_{GET} =
	{1\over2}(\rho |v|^2 + p^{ii} - \Upsilon \Xi^{-1} \rho)
\]

\subsection{Energy-momentum tensor}

Now, we have derived the GET Lagrangian using assumptions about the
conservation laws.  Therefore, to write dowm the energy-momentum
tensor is easy:

\[T^\mu_\nu = (4\pi G)^{-1}\gamma_{\nu\kappa}g^{\kappa\mu}\sqrt{-g}\]

In this form, the tensor does not depend on the material properties
$\varphi^m$ of the ether.  But how is this energy-momentum tensor
related with the usual energy-momentum tensor for the matter fields?
Now, the answer is simple.  We have to multiply the GET variant of the
Einstein equation (\ref{GET-Einstein}) with $\sqrt{-g}$ and obtain the
following decomposition of the full energy-momentum tensor:

\[
T^\mu_\nu =
  (T_m)^\mu_\nu\sqrt{-g}
  + (8\pi G)^{-1}\left(
  	(\Lambda +\gamma_{\kappa\lambda}g^{\kappa\lambda})\delta^\mu_\nu
  	- G^\mu_\nu
  \right)\sqrt{-g}
\]

Thus, we obtain immediately what is missed in GR: an energy-momentum
tensor for the gravitational field.

\subsection{Constraints} \label{constraints}

If we want to formulate an initial value problem for GR as well as
GET, we cannot simply define the initial values
$g^0_{\mu\nu}(x)=g_{\mu\nu}(x,0)$ and
$k^0_{\mu\nu}(x)=\partial_t g_{\mu\nu}(x,t)|_{t=0}$.
Instead, we obtain the problem that the four equations

\[ G^0_\mu = ... \]

do not contain second order derivatives in time, that means, they
define constraints for the initial values.  This is a common property
in above theories, because the additional terms of GET do not add
second order derivatives of the $g_{\mu\nu}$.  Moreover, in GET
the four conservation laws are also only first order in time.

Nonetheless, their character is completely different. In GR, these
constraints play a very special role in the ADM Hamilton formalism --
the energy H itself is a constraint.  This is a consequence of the
covariance of the GR Lagrangian.  In GET, as we have already seen, we
have a well-defined local energy and momentum density.

Even if the constraints are much more harmless in GET, they remain to
be constraints, which is not very nice.  But there is some interesting
insight into the nature of the constraints in GET which has been found
for GR in harmonic coordinates by Choquet-Bruhat
\cite{Choquet-Bruhat}.  This insight was important for his proof of
local existence and uniqueness theorems for GR.  It is remarkable in
itself that this proof has been done in harmonic coordinates.

First, as has been observed by Lanczos \cite{Lanczos}, the Ricci
tensor essentially simplifies in harmonic coordinates:

\[ R^{(h)}_{\mu\nu} = -{1\over2}g^{\alpha\beta}
    {\partial^2 g_{\mu\nu}\over\partial x^\alpha\partial x^\beta}
    + H_{\mu\nu}
\]

where $H_{\mu\nu}$ does not contain second derivatives of the metric.
Now, in GET we have the harmonic condition
$\Gamma^\mu=\partial_\nu(g^{\mu\nu}\sqrt{-g})=0$ as an equation.
Therefore, for the initial values we have

\[ \Gamma^\mu(0,x) = 0 \, \hspace{1cm} \partial_t \Gamma^\mu(0,x) = 0\]

The second condition contains a second order time derivative.  But
this is true also for the Ricci tensor in harmonic coordinates.  Now,
if we use the appropriate combination of this second initial condition
and the equation in harmonic coordinates, we obtain the four other
first order constraints (see \cite{Fisher}, Lemma 22).  Thus, all
constraints are closely related to the harmonic condition.

The ether interpretation gives additional insight.  The point is that
in this interpretation the components $\rho v^i = g^{0i}\sqrt{-g}$ are
already velocities.  A second order equation for these components
would be a third order equation for the ether particles them-self.
Therefore, it is very natural that there are no such third order terms
in the equations them-self.

\section{Derivation of the Einstein equivalence principle}

The Einstein equivalence principle (EEP) is an immediate consequence
of the GET Lagrange density: the matter Lagrangian is covariant in the
strong sense, does not depend on the preferred coordinates $X^\mu$.
The question we want to consider here is if there are generalizations
of the GET axioms so that the EEP remains correct.

At first we have to formulate the EEP in an appropriate way.  The
equations for matter do not depend on the preferred coordinates.  In
our covariant formalism there is a natural way to do formulate this
property.  The ``equations for the matter'' we identify with the
Euler-Lagrange equations for $\varphi^m$, and the property that they
do not depend on the preferred coordinates $X^\mu$ can be written as

\[ {\delta\over\delta X^\mu} {\delta S\over\delta\varphi^m} = 0\]

To obtain a proof, let's look how this property may be proven for the
GET Lagrangian:

\begin{theorem}[Einstein equivalence principle]
Let L be a weak covariant Lagrangian with the conservation laws

\[ \partial_{\mu}T^\mu_\nu = 0. \]

If the conservation laws do not depend on the variables $\varphi^m$,
then the Einstein equivalence principle holds for these variables.

\end{theorem}

Proof: The conservation laws in the weak covariant formalism are
defined as

\[ {\delta S\over\delta X^\mu} = \partial_{\nu}T^\nu_\mu = 0 \]

The EEP follows immediately:

\[
   {\delta\over\delta X^\mu}    {\delta S\over\delta\varphi^m}
 = {\delta\over\delta\varphi^m} {\delta S\over\delta X^\mu}
 = {\delta\over\delta\varphi^m} \partial_{\nu}T^\nu_\mu
 = 0
\]

As we see, the first property we need is that the material properties
are not used in the conservation laws.  This property depends on our
choice of $p^{ij}$ as an independent variable.  But this is only a
technical question.  More important is the universality axiom -- that
the matter fields $\varphi^m$ describe only ``material properties'' of
the ether.  External steps of freedom, that means other, non-ether
fields which interact with the ether have some momentum exchange with
the ether.  Therefore, there will be some interaction terms in the
momentum conservation laws.

But this may be partially weakened.  If some steps of freedom $\psi^n$
are external forces or external matter, while other steps of freedom
$\varphi^m$ describe material properties of the ether, than the EEP
does not hold for the external steps of freedom, but remains valid for
the material properties.  The proof remains the same.

The other property is that the conservation laws are the
Euler-Lagrange equations for the preferred coordinates:

\[ {\delta S\over\delta X^\mu} =  \partial_{\nu}T^\nu_\mu \]

Now, this may be generalized for the case where we have explicit
dependencies on the preferred coordinates, and, therefore, no
conservation laws.  All we need is that the equation does not depend
on the material properties $\varphi^m$.

Therefore, we conclude that the EEP holds for material properties
$\varphi^m$ even in more general situations, if we have other external
fields, external forces, even explicit dependencies of the Lagrangian
from the coordinates.  Let's summarize: {\em The EEP holds for a step
of freedom $\varphi^m$ only if it describes a material property of the
ether.  The universality axiom explains why the EEP holds for all
matter fields.\/}

\subsection{Higher order approximations in a Lagrange formalism}

Let's consider now another possibility for generalization.  We
consider the situation where we have to consider different
approximations for a Lagrange formalism.  All we assume is that all
approximations are consistent.  For a condensed matter theory that
means that the conservation laws are valid.  Let's compare now two
approximations.  For the approximation $S = S_0 + S_1$ we have

\[{\delta S_0 \over\delta X^\mu} = \partial_{\nu}T^\nu_\mu \]

as well as

\[{\delta S \over\delta X^\mu} = \partial_{\nu}T^\nu_\mu \]

Now, in this situation we do not even need that these conservation
laws do not depend on some fields $\varphi^m$.  All we need now is
that the two approximations of the conservation laws are identical.
This leads immediately to the Einstein equivalence principle for the
additional part of the Lagrangian:

\[{\delta S_1 \over\delta X^\mu} = 0 \]

This consideration suggests that it may be even easier to detect
relativistic symmetry in the higher order approximations.  Sequences
$S_n$ of approximations appear in effective field theory.

\subsection{Weakening the assumptions about the Lagrange formalism}

We have obtained the conservation laws with reference to condensed
matter theory.  Then we have identified them with the conservation
laws from the weak covariant formalism.  This is a quite natural, but
non-trivial identification.  There are many different variants of the
conservation laws, and even if they are equivalent if the equations of
motion are fulfilled, their functional dependencies differ.
Therefore, it would be nice to weaken this assumption.

Unfortunately, it seems impossible to prove something without an
explicit assumption which relates a certain conservation law with the
Euler-Lagrange equations.  The property ``there exists a Lagrange
formalism'' for some equivalent system of equations is too weak.  The
problem is that there are various methods of transformation of a given
system of equations -- multiplying them with ``integrating factors'',
Lagrange multipliers, replacement of fields by potentials.  That's why
a general method which allows to decide if a given set of equations is
equivalent to a system of Euler-Lagrange equations is not known
\cite{Wagner}.  That means, we are not even able to find all Lagrange
formalisms for a given set of equations.

It seems natural to assume that the equations already have the form of
Euler-Lagrange equations.  Such systems of equations are
``self-adjoint'' and have been considered in detail \cite{Santilli}.
Especially there exist standard methods to construct Lagrange
densities which are also tests if the system is self-adjoint
\cite{Wagner}.  Thus, let's assume that the conservation laws
are part of such a self-adjoint system of equations, that means, are
Euler-Lagrange equations for some variables $c^\mu$:

\[ {\delta S\over\delta c^\mu} = \partial_\nu T^\nu_\mu \]

This allows to derive a similar symmetry property of the equations for
$\varphi^m$:

\[
   {\delta \over\delta c^\mu} {\delta S\over\delta \varphi^m}
 = {\delta \over\delta \varphi^m}{\delta S\over\delta c^\mu}
 = {\delta \over\delta \varphi^m}\partial_\nu T^\nu_\mu
 = 0
\]

Thus, we have a symmetry group with four continuous parameters, only
the relation between these parameters and the coordinates has been
lost.

\subsection{Explanatory power of the derivation}

Last not least, it should be noted that the derivation of the EEP has
very high explanatory power.  First, as we have seen, it is based on a
few very general principles.  We do not need any special assumptions
about the ether, no ``mechanical explanation'', no strange
``mechanism'', no ``conspiracy''.  We make non-trivial assumptions,
but these non-trivial assumptions are very natural for a condensed
matter theory.

Evidence for the high explanatory power is that we can describe the
proof in a simple verbal way: The conservation laws can be understood
as equations for the preferred coordinates.  Now, the conservation
laws do not depend on the material properties.  Therefore, because of
the principle ``action equals reaction'', the equations for the
material properties do not depend on the preferred coordinates.

\section{Does usual matter fit into the GET scheme?}

If we compare the Einstein equations with usual hydro-dynamical
equations, they look very different.  The Einstein equations depend on
second order derivatives of the $g_{\mu\nu}$.  At a first look, there
seems to be no chance to unify them.

GET suggests such a way.  Indeed, in the derivation of GET we have
used only a few general properties of the ether.  None of these
assumptions is obviously wrong for usual condensed matter theory.  The
material properties of the GET ether remain unspecified.  We cannot
even tell if the ether is solid or liquid.\footnote{It may be assumed
that the condition that the pressure tensor $p^{ij}$ is negative
definite tells something about the ether -- that the ether is a
material with negative pressure.  But the only reason for naming the
tensor field $p^{ij}$ ``pressure'' is that it appears like usual
pressure appears in the usual Euler equation.  And this allows to
identify $p^{ij}$ with usual pressure only modulo a constant.}  This
suggests that usual condensed matter may be described by a GET-like
Lagrangian.

\subsection{The role of the Einstein Lagrangian}

The problem with the second derivatives which appear in the Einstein
equations can be easily solved: we have identified the ``remaining
part'' of the GET Lagrangian with the Einstein Lagrangian because of
its strong covariance.  None of the GET axioms requires to include the
Einstein-Hilbert term $L_{GR} = R \sqrt{-g}$ into the GET Lagrangian.
It is simply a possible term in the GET Lagrangian, not a necessary
one.  The same holds for covariant terms with higher
order.\footnote{As described by Weinberg \cite{Weinberg} ``there's no
reason in the world to suppose that the Lagrangian does not contain
all the higher terms with more factors of the curvature and/or more
derivatives, all of which are suppressed''.}  Moreover, in comparison
with the three ``cosmological terms''
$g^{00}\sqrt{-g},g^{ii}\sqrt{-g},\sqrt{-g}$ of GET the
Einstein-Hilbert term is a higher order term.  Therefore, in the first
approximation for usual condensed matter the Einstein-Hilbert term may
be simply omitted.  In this case, the GET equation becomes an
algebraic relation between the gravitational field defined by
$\rho,v^i,p^{ij}$ and the material properties $\varphi^m$.  This is
already much more close to usual condensed matter theory.

On the other hand, GET suggests that the Einstein-Hilbert Lagrangian
should be used in higher order approximations of condensed matter
theory.  The physical meaning of curvature-dependent term in condensed
matter theory is easy to understand: if curvature is zero, then there
exists an undistorted reference state which remains unchanged in time.
Therefore, curvature describes inner stress and its change in time.

\subsection{The role of Lorentz symmetry}

Another property seems to be much more in contradiction with usual
hydrodynamics -- the Lorentz invariance of the GET Lagrangian.  The
physical meaning of this Lorentz invariance is not clear.

One possibility is that it has none, and it simply as a consequence of
the fact that there are not very much possibilities for quadratic
Lagrangians, therefore, symmetries appear more or less by accident.
With another choice of the constants, especially $Xi<0$, we would
obtain an SO(4) symmetry, also without any physical meaning.

Nonetheless, a consequence is that we cannot describe with GET a
Galilean-invariant theory.  On the other hand, we can use GET to
describe special-relativistic condensed matter.  The reverse method
would be to describe a Galilean invariant theory as the limit $\Xi\to
0$ of a GET theory.

\subsection{Existing research about the similarity}
\label{existing_ether_research}

Considering the mentioned problem to obtain a Galilean invariant
theory it is no wonder that the usual Lagrange formalisms for
non-relativistic fluid dynamics in Euler coordinates (as far as
considered in Wagner \cite{Wagner}) do not fit into our scheme.
Remarkably, to use the three-dimensional Einstein-Hilbert Lagrangian
to describe dislocations has been proposed by Malyshev
\cite{Malyshev}.

On the other hand, it is widely acknowledged in the condensed matter
community that phonons in various matter move in an effective Lorentz
metric $g_{\mu\nu}$ which is usually curved.  Various aspects have
been considered here.  Katanaev and Volovich \cite{Katanaev} compare
wedge dislocation with cosmic strings. See also Guenther
\cite{Guenther}.  A lot of research has been related with the idea of
``dumb holes'' -- an analog of ``black holes'' in acoustics.  These
``dumb holes'' may be used to study quantum gravity effects like Unruh
radiation in usual condensed matter (Unruh \cite{UnruhAcoustic},
Jacobson \cite{Jacobson91}, Rosu
\cite{Rosu}, Visser \cite{VisserAcoustic}).  

The most interesting example of usual condensed matter is superfluid
$^3He$.  Here not only a curved Lorentz metric has been identified,
but also chiral fermions and non-abelian gauge fields (Volovik
\cite{Volovik}, Jacobson and Volovik \cite{Jacobson98}).

Of course, the amount of research which connects condensed matter
theory and fundamental field theory is much greater.  As noted by
Wilczek \cite{Wilczek}, ``the continuing interchange of ideas between
condensed matter and high energy theory, through the medium of quantum
field theory, is a remarkable phenomenon in itself.  A partial list of
historically important examples includes global and local spontaneous
symmetry breaking, the renormalization group, effective field theory,
solitons, instantons, and fractional charge and statistics.''

\section{Comparison with RTG} \label{RTG}

There is also another theory with almost the same Lagrangian -- the
``relativistic theory of gravity'' (RTG) proposed by Logunov et
al. \cite{Logunov}.  In this theory, we have a Minkowski background
metric $\eta_{\mu\nu}$.  The Lagrangian of RTG is

\[L = L_{Rosen} + L_{matter}(g_{\mu\nu},\psi^m)
    - {m_g^2\over 16\pi}({1\over 2}\eta_{\mu\nu}g^{\mu\nu}\sqrt{-g}
            - \sqrt{-g} - \sqrt{-\eta})
\]

If we identify the Minkowski coordinates in RTG with the preferred
coordinates in GET, the Lagrangians are equivalent as functions of
$g_{\mu\nu}$ for the following choice of constants:
	$\Lambda=-{m_g^2\over 2}<0$,
       	$\Xi=-\eta^{11}{m_g^2\over 2}>0$,
	$\Upsilon=\eta^{00}{m_g^2\over 2}>0$.
In this case, the equations for $g^{\mu\nu}$ coincide.  The harmonic
equation for the Minkowski coordinates hold in RTG \cite{Logunov}.  As
a consequence, the equations of the theories coincide.

Nonetheless, the Euler-Lagrange equations are not all.  In above
theories we have additional restrictions related with the notion of
causality -- causality conditions.  In GET, causality is related with
the Newtonian background -- the preferred time $T(x)$ should be a
time-like function.  This is equivalent to the condition $\rho>0$.  In
RTG, causality is defined by the Minkowski background.  The light cone
of the physical metric $g_{\mu\nu}$ should be inside the light cone of
the background metric $\eta_{\mu\nu}$.

Once RTG is a special-relativistic theory, it is also incompatible
with the EPR criterion of reality and Bohmian mechanics.  This
question should be considered as the most serious difference.

There is also a difference in the quantization concept.  RTG suggests
to apply standard quantum field theory on a Minkowski background,
while GET suggests to understand quantum field theory as an effective
field theory.  The GET prediction about the cutoff length depends on
the interpretation of $g^{00}\sqrt{-g}$ as the density of the ether
and is not Lorentz-invariant.

RTG has a completely different metaphysical background.  Therefore,
RTG has a completely different justification of the Lagrangian.  While
such metaphysical differences are often considered to be unimportant
in physics, we do not agree.  Metaphysical interpretations and
esthetic feelings often influence preferences for theories.  Because
the simplicity and beauty of the explanation of the Einstein
equivalence principle is one of the main advantages of GET, this
question should not be underestimated.

\section{Comparison with GR with four dark matter fields} \label{GRDM}

There is also another theory with the same Lagrangian -- GR with four
scalar ``dark matter fields'' $X^\mu(x)$\footnote{Kuchar \cite{Kuchar}
has considered similar scalar fields in GR as ``clock fields''.}.
Let's denote it as GRDM.  The Lagrangian is

\[L_{GRDM} = -(8\pi G)^{-1}\gamma_{\mu\nu}X^\mu_{,\alpha}X^\nu_{,\beta}
  	g^{\alpha\beta}\sqrt{-g}
  + L_{GR}(g_{\mu\nu}) + L_{matter}(g_{\mu\nu},\varphi^m) \]

and therefore formally equivalent to GET.  But we have completely
forgotten the physical meaning of the fields $X^\mu(x)$ as
coordinates.  In GRDM, they are really only scalar fields.

We have to consider here the usual GR energy conditions.  They require
the following signs: $\Upsilon<0$ and $\Xi>0$.

There is another difference between GET and GRDM which is essential
and important to understand.  In GET, we have additional global
restrictions:

\begin{itemize}

\item  First, the fields $X^\mu(x)$ are global coordinates in GET.
In GRDM, there will be many solutions where the dark matter fields do
not define a global system of coordinates.  Moreover, it will be even
the typical solution of GRDM.  Indeed, solutions which define global
coordinates have unusual boundary conditions.  Moreover, complete
classes of solutions are excluded: all solutions with non-trivial
topology are forbidden.

 \item Second, the coordinate $T(x)=X^0(x)$ is a global time-like
function.  This is equivalent to $\rho>0$.  Again, a whole class of
solutions of GRDM is excluded: all solutions with closed time-like
curves.

 \item We have also other, unusual boundary conditions for the fields
$X^\mu(x)$: their boundary values go to infinity.

\end{itemize}

These properties do not follow from the Euler-Lagrange equations.
Instead, we have to remember that the original axioms are axioms about
an ether in a Newtonian space-time.  The Lagrange formalism with the
``fields'' $X^\mu(x)$ is only derived, not fundamental.

Therefore, it is in no way a weakness of GET that the additional
global restrictions do not follow from the GET equations.  Instead,
this proves additional predictive power of GET in comparison with
GRDM.  Indeed, if we observe a solution of GRDM which does not fulfill
the additional global restrictions we have falsified GET, but not
GRDM.  Thus, in GET we have additional possibilities for
falsification, therefore, higher empirical content.

This difference is of principal, conceptual character -- similar
global restrictions are impossible in general-relativistic theories.
Therefore, it helps to understand the difference between
general-relativistic theories and theories with preferred frame.
We consider this question also in appendix \ref{relationalism}.

\section{Comparison with General Relativity}

Let's consider now the differences between the predictions of GET and
GR itself.  There are, first, the differences between GET and the
variant of GR with four dark matter fields we have already considered.
This restricts GET to global hyperbolic solutions with trivial
topology, moreover, of a special type -- with global harmonic
coordinates and global harmonic time-like function.

\subsection{Dark matter and energy conditions}

The other part of the difference between GET and GR can be understood
as the difference between GRDM and GR.  First, the additional terms
define ``dark matter'' in the sense that the scalar fields $X^\mu(x)$
do not interact with usual matter.

To understand the most interesting property of this new type of ``dark
matter'' we have to consider the energy conditions.  GET does not yet
relate energy conditions with general properties of the ether.  But
because GR does not provide an explanation too, this is unproblematic.
We can introduce them -- as in GR -- as additional, yet unexplained,
properties of the matter Lagrangian.  But if we introduce it in this
way, as a property of the matter fields, the energy conditions do not
restrict the sign of $\Xi$ and $\Upsilon$.  Therefore, without
violating the GET version of the energy condition, it is possible to
set $\Upsilon>0$.  And, because of the interesting predictions which
follow from this choice, we really set $\Upsilon>0$.

From point of view of GRDM, this choice violates all energy
conditions.  Therefore, GRDM contains a very special, strange type of
``dark matter'' which violates all energy conditions of GR.
Therefore, all general theorems about GR which use various energy
conditions fail.  Especially the theorems about big bang and black
hole singularities fail.

\subsection{Homogeneous universe: no big bang singularity}
\label{Friedman}

Let's consider at first the homogeneous universe solutions of the
theory.  Because of the Newtonian background frame, only a flat
universe may be homogeneous. Thus, we make the ansatz:

\[ ds^2 = d\tau^2 - a^2(\tau)(dx^2+dy^2+dz^2) \]

Now, we see that in this ansatz the spatial coordinates $x^i$ are
already harmonic.  It remains to find the harmonic time.  The equation
for harmonic time is $dT/d\tau=1/a^3$.  The metric in harmonic
coordinates is therefore:

\[ ds^2 = a^6(t) dT^2 - a^2(t) \delta_{ij}dX^i dX^j \]

Note that $\rho=g^{00}\sqrt{-g}=1$, thus, in this ansatz the ether has
constant density, and the universe does not expand.  The observable
expansion is an effect of shrinking rulers.  In GR this would be only
one of the possible interpretations, without physical importance.
Instead, in GET this is the preferred interpretation because of
symmetry reasons.  Thus, if the universe is homogeneous, the universe
does not expand, but our rulers are shrinking.  But, as in GR, the
global universe looks like expanding.

Below we use standard relativistic language and usual proper time
$\tau$ (that means,$\dot{a}=\partial a/\partial\tau$). Using some
matter with $p=k\varepsilon$ we obtain the equations ($8\pi G = c =
1$):

\begin{eqnarray*}
3(\dot{a}/a)^2  &=&
  	- \Upsilon/a^6 + 3 \Xi/a^2 + \Lambda + \varepsilon\\
2(\ddot{a}/a) + (\dot{a}/a)^2 &=&
  	+ \Upsilon/a^6 +   \Xi/a^2 + \Lambda - k \varepsilon
\end{eqnarray*}

The $\Upsilon$-term influences only the early universe, its influence
on later universe may be ignored.  But, if we assume $\Upsilon >0$,
the qualitative behavior of the early universe changes in a
remarkable way.  We obtain a lower bound $a_0$ for $a(\tau)$ defined
by

\[ \Upsilon/a_0^6 = 3 \Xi/a_0^2 + \Lambda + \varepsilon \]

The solution becomes symmetrical in time.  Therefore, before the big
bang there was a big crush, and the whole story can be named big
bounce.  For some simple situations, analytical solutions are
possible.  For example, if $\varepsilon = \Xi =0, \Upsilon>0,
\Lambda>0$ we have the solution

\[ a(\tau) = a_0 \cosh^{1/3}(\sqrt{3\Lambda} \tau)  \]

\subsection{Is there independent evidence for inflation theory?}

Now, such a big bounce scenario solves the problems of the big bang
scenario with the small horizon.  For the description of these
problems and their current solution in inflation theory we follow
Primack \cite{Primack}.  There are two such problems with a small
horizon: First, ``the angular size today of the causally connected
regions at recombination ($p^+ + e^-\to H$) is only $\Delta\theta\sim
3^o$.  Yet the fluctuation in the temperature of the cosmic background
radiation from different regions is very small: $\Delta T/T\sim
10^{-5}$. How could regions far out of causal contact have come to
temperatures that are so precisely equal?  This is the `horizon
problem'.''  (p.56)

Even more serious seems the following problem: In the standard hot big
bang picture, ``the matter that comprises a typical galaxy, for
example, first came into causal contact about a year after the big
bang.  It is hard to see how galaxy-size fluctuations could have
formed after that, but even harder to see how they could have formed
earlier'' (p.8).

Last not least, there is the ``flatness problem''.  In GR, the
assumption that the universe is flat does not seem to be natural.  But
for a curved universe, the initial curvature has to be extremely small
in comparison to a natural dimensionless constant for curvature.

Now, these three problems seem sufficient to rule out the standard Big
Bang model without inflation.  But all three problems are solved in
GET without inflation: the two variants of the horizon problem are
solved because the horizon in a universe with big bounce is much
larger, if not infinite.  And the flat universe is certainly preferred
as the only homogeneous universe, therefore, there is no flatness
problem.

Therefore, it seems reasonable to question the necessity for inflation
in GET cosmology.  There are some other problems solved by inflation:
that it ``dilutes any preceding density of monopoles or other
unwanted relics'', and predictions about a ``nearly constant curvature
spectrum $\delta_H =$ constant of adiabatic fluctuations'' (p.59).  If
these will be serious problems for a GET universe without inflation is
hard to say.  If such ``relics'' are really necessary because of
particle theoretical reasons, it seems possible to use a large enough
$\Upsilon$.  Then the critical temperature which causes the creation
of the various ``relics'' may not be reached during the Big Bounce.
What GET allows to predict about the spectrum of adiabatic fluctuation
will be a question for future research, but it seems not unreasonable
to assume that the simplest imaginable spectrum -- the constant one --
may be compatible with GET.

Another question is if inflation is a necessary consequence of
particle theory.  This seems to be not the case.  To obtain inflation,
we have to make non-trivial assumptions about this phase
transition.\footnote{``In the first inflationary models, the dynamics
of the very early universe was typically controlled by the self-energy
of the Higgs field associated with the breaking of a Grand Unified
Theory (GUT) into the standard 3-2-1 model: $GUT\to
SU(3)_{color}\otimes\left[SU(2)\otimes U(1)\right]_{electroweak}$.
... Guth (1981) initially considered a scheme in which inflation
occurs while the universe is trapped in an unstable state (with the
GUT unbroken) on the wrong side of a maximum in the Higgs potential.
This turns out not to work ... The solution in the `new inflation'
scheme ... is for inflation to occur after barrier penetration (if
any). It is necessary that the potential of the scalar field
controlling inflation (`inflaton') be nearly flat (i.e. decrease very
slowly with increasing inflaton field) for the inflationary period to
last long enough.  This nearly flat part of the potential must then be
followed by a very steep minimum, in order that the energy contained
in the Higgs potential be rapidly shared with the other degrees of
freedom (`reheating').  A more general approach, `chaotic' inflation,
has been worked out ... However, ... it is necessary that the inflaton
self-coupling be very small ... This requirement prevents the Higgs
field from being the inflaton.'' \cite{Primack}, p. 57. This
consideration suggests that inflation is in no way a necessary
consequence of the phase transition related with a GUT.}  Particle
theory does not give independent evidence in favor of inflation.
Thus, it seems that GET cosmology with $\Upsilon>0$ is a viable theory
without inflation, while GR requires inflation.

The existing evidence for a hot state of the universe may be used to
obtain upper bounds for $\Upsilon$.

\subsection{A new dark matter term}

The influence of the $\Xi $-term on the age of the universe is easy to
understand.  For $\Xi >0$ it behaves like homogeneously distributed
dark matter with $p=-(1/3)\varepsilon$.  It influences the age of the
universe.  A similar influence on the age of the universe has a
non-zero curvature in GR cosmology.

It seems not unreasonable that a non-zero value for $\Xi $ may be part
of the solution of the dark matter problem.  According to Primack
\cite{Primack}, there seems to be a large amount of ``cold dark
matter'' (CDM), but this is not sufficient to fit the data.  The
models favored in this paper have some additional homogeneous
component: some ``hot dark matter'' part (CHDM) or a non-zero
cosmological constant ($\Lambda$CDM).  Now, the ``dark matter'' term
proposed here is also homogeneous, and something between homogeneous
``hot dark matter'' and a cosmological constant.

Thus, the $\Xi$-term defines a reasonable candidate for dark
matter.   Current observation seems to favor $\Xi>0$.

\subsection{Stable frozen stars instead of black holes}
\label{frozen_stars}

Let's consider now spherical symmetric stable solutions.  Of course,
for symmetry reasons, we want to have static preferred coordinates
too.  For the preferred coordinates $X^i$ the metric should be
harmonic.  Of course, we describe this metric using the ``preferred
radius'' $r = \sqrt{\delta_{ij}X^iX^j}$.  Fortunately, there is a
simple general formula for the harmonic metric.  For a given function
$m(r), 0<m<r$, the metric

\[ ds^2 = (1-{m \partial m/\partial r \over r})
      	     	({r-m\over r+m}dt^2-{r+m\over r-m}dr^2)
       	- (r+m)^2 d\Omega^2 \]

is harmonic in $X^i$.  For constant m, this formula reduces to the
Schwarzschild metric in harmonic coordinates.  Therefore, m(r) defines
the (harmonic) Schwarzschild radius of the mass inside the radius r.

Now this general solution of the harmonic equation may be used to
construct various partial solutions for special matter equations.  We
start with an arbitrary distribution of mass m(r) with $0<m<r,
\partial m/\partial r\ge 0$.  The Einstein equations define
$\varepsilon(r)$ and $p(r)$, and we obtain a solution for some
material law which depends on the radius: $p = k(r)
\varepsilon$.  As a simple example, let's consider the ansatz
$m(r)=(1-\Delta)r$.  We obtain

\begin{eqnarray*}
ds^2 &=& \Delta^2dt^2 - (2-\Delta)^2(dr^2+r^2d\Omega^2) \\
0    &=& -\Upsilon \Delta^{-2} +3\Xi(2-\Delta)^{-2}+\Lambda+\varepsilon\\
0    &=& +\Upsilon \Delta^{-2} + \Xi(2-\Delta)^{-2}+\Lambda-p
\end{eqnarray*}

Now, in GR we obtain only the trivial solution $\varepsilon = p = 0$.
Once the cosmological constants are sufficiently small, nothing
changes for moderate values of $\Delta$. But for sufficiently small
$\Delta\ll 1$ the situation changes -- we obtain a stable solution $p
= \varepsilon = \Upsilon \Delta^{-2}$.  This is a stable star with a
radius very close to the Schwarzschild radius, with time dilation
$\Delta^{-1}=\sqrt{\varepsilon/\Upsilon} \sim M^{-1}$ for a frozen
star of mass M.

It is seems obvious that this is not a special property of this
solution, but a rather general effect.  Even a very small
$\Upsilon$-term becomes important close enough to the horizon size and
allows to obtain stable solutions.  For a collapsing star this term
defines a counter-force which stops the collapse immediately before
horizon formation and leads to a subsequent explosion.  This explosion
does not follow immediately, because near the bounce the movement is
time-dilated too.

Thus, we obtain very interesting differences for the gravitational
collapse.  For the outside observer, we can fit the GR predictions
making $\Upsilon$ small enough.  But even for arbitrary small
$\Upsilon>0$ we have remarkable qualitative differences -- there is no
region ``behind the horizon'', no singularity, and every infalling
observer can observe this difference.

\section{General-relativistic quantization problems}

The quantization of gravity is usually considered as one of the major
problems of fundamental physics.  But, it seems, this problem should
be named instead ``quantization of general relativity''.  Indeed,
Butterfield and Isham \cite{Butterfield} note that ``... most workers
would agree on the following ... diagnosis of what is at the root of
most of the conceptual problems of quantum gravity. Namely: general
relativity is not just a theory of the gravitational field -- in an
appropriate sense, it is also a theory of spacetime itself; and hence
a theory of quantum gravity must have something to say about the
quantum nature of space and time.''

Now, in GET the theory of gravity is not a theory of space-time
itself, instead, it is a theory of a medium in a classical Newtonian
space-time.  This problem is, obviously, much easier -- the most
serious problems simply disappear.  On the other hand, it is much less
interesting -- we do not learn anything new about ``the quantum nature
of space and time'', instead, the classical Newtonian space-time
defines the fixed stage for quantization.  While the Newtonian
background is very simple, it is also not very interesting and remains
as unexplained as in Newton's theory.  The really hard, conceptual,
interesting problems of relativistic quantum gravity disappear into
nothing.  What remains seem to be only a few technical problems as
complex as the quantization of usual condensed matter.

Once these conceptual problems disappear, they can be considered as
additional fundamental support for GET and are therefore worth to be
considered in this context.  In appendix \ref{superposition} we
consider in more detail a problem related with superposition of
gravitational fields which I have named the ``scalar product
problem''.  This problem has several nice properties: it suggests a
simple solution -- a fixed space-time background which is common for
different gravitational fields.  Moreover, it is based on an
interesting quantum observable -- a transition probability.  And this
transition probability may be computed in the non-relativistic limit
-- multi-particle Schr\"odinger theory.

The notorious ``problem of time'' is mainly a conceptual problem.  It
appears if we make a deliberate theoretical decision: that the time
measured with clocks -- the time of general relativity -- has to be
unified with the notion of time of quantum mechanics.  We discuss
these metaphysical questions in \S{ } \ref{time}.

Nonetheless, some other well-known problems of quantum GR which do not
exist in quantum GET are also worth to be considered: the problem of
causality, and the information loss problem.

\subsection{Causality}

The problem of time is also closely related with causality: ``General
relativity accustoms us to the ideas that (i) the causal structure of
spacetime depends on the metric ... and (ii) the metric and causal
structure are influenced by matter ... In general relativity, these
ideas are `kept under control' in the sense that in each model, there
is of course a single metric tensor $g_{\mu\nu}$, representing a
single metric and causal structure.  But once we embark on
constructing a quantum theory of gravity, we expect some sort of
quantum fluctuations in the metric, and so also in the causal
structure.  But in this case, how are we to formulate a quantum theory
with a fluctuating causal structure?'' \cite{Butterfield}

This conceptual problem has also technical aspects.  ``For example, a
quantum scalar field satisfies the micro-causal commutation relations

\[ \left[\hat{\phi}(X),\hat{\phi}(Y)\right] = 0 \]

whereby fields evaluated at space-like separated spacetime points
commute.  However, the concept of two points being space-like separated
has no meaning if the spacetime metric is probabilistic or
phenomenological.  In the former case, the most likely scenario is
that [the commutator] never vanishes, thereby removing one of the
foundations of conventional quantum field theory.''

\subsection{Information loss problem}

Another problem which disappears is the ``information loss problem''
proposed by Hawking \cite{Hawking76}.  The problem is that the black
hole contains information.  But the Hawking radiation cannot take away
this information because it is determined only by the geometry of the
black hole outside the horizon, and the black hole has no hair that
records any detailed information about the collapsing body.  The key
constraint comes from causality -- once the collapsing body is behind
the horizon, it is incapable of influencing the radiation.  Now,
suppose the black hole evaporates.  That means, the black hole has
been replaced by the radiation completely.  It is a familiar fact of
life that information is often lost in practice.  But here the
information is lost in principle.  It seems that an initially pure
state becomes after evaporation a mixed state.  And this is in
contradiction with the fundamental principles of quantum
mechanics.

Preskill \cite{Preskill} in a review of the problem writes that
initially he ``was inclined to dismiss Hawking's proposal as an
unwarranted extrapolation from an untrustworthy approximation''. But
as ``I have pondered this puzzle, it has come to seem less and less
likely to me that the accepted principles of quantum mechanics and
relativity can be reconciled with the phenomenon of black hole
evaporation.''

Now, it is hard for me to judge about the seriousness of this problem.
Personally I'm convinced that the correct black hole evaporation
scenario in GR is different from the usually accepted one.  In my
opinion it is the scenario of black hole evaporation proposed by
Gerlach \cite{Gerlach}.  In this scenario, no black hole horizon is
formed.  As far as I understand, relativists do not like this scenario
because it seems to prefer the coordinates of the outside observer.
But I don't think this argument is justified -- the preference is
predefined by the preference for the Minkowski vacuum state in the
initial situation before the collapse.  In Gerlach's scenario we have
no information loss problem because no horizon is formed.

Anyway, in GET the information loss problem disappears together with
the black holes.  We have stable frozen stars which do not radiate
Hawking radiation once they have reached a stable state.  We also have
not to be afraid of some similar situations -- there is always a
global absolute time, and this global time has to be used in quantum
GET to define an unitary evolution.

\section{Atomic ether theory}

With the Newtonian background only the conceptual problems related
with relativistic space-time quantization disappear.  The problem with
non-re\-normali\-za\-bility remains.  But there is a natural solution
for this problem in the context of a condensed matter theory -- an
atomic hypothesis.  Therefore, we assume that our medium has an atomic
structure.  This leads to an explicit cutoff.  The regularization
becomes physical.  The concept of gravity as an effective field theory
is well-known and goes back to Sakharov \cite{Sakharov}.
\footnote{Jegerlehner notes that ideas that ``the relationship between
bare and renormalizes parameters obtains a physical meaning ... are
quite old and in some aspects are now commonly accepted among particle
physicists'' \cite{Jegerlehner}.  Weinberg describes this as ``the
present educated view on the standard model, and of general
relativity, ... that these are leading terms in effective field
theories'' \cite{Weinberg}.}

One property of this widely accepted effective field theory picture is
that it makes a certain assumption about the cutoff length.  It is
assumed to be the Planck length $a_P\simeq 10^{-33}$cm.  This property
has even used to name this concept: ``Planck ether'', ``Planck solid''
\cite{Jegerlehner} or ``Planck condensed matter'' \cite{Volovik}.
But this is de facto the only property of the ``Planck ether'' which
is known.
\footnote{``The curvature of
space-time is relevant and special relativity is modified by
gravitational effects.  One expects a world which exhibits an
intrinsic cutoff corresponding to the fundamental length $a_P\simeq
10^{-33}$cm. But not only Poincare invariance may break down, also the
laws of quantum mechanics need not hold any longer at $\Lambda_P$.''
\cite{Jegerlehner}}

The ``atomic ether hypothesis'' differs from this concept in
everything except the fact that current field theory should be
replaced by another one below some cutoff.  First, a well-defined
space-time concept has been fixed -- the classical Newtonian
background with Euclidean space, absolute time and classical
causality.  Moreover we have well-defined conservation laws.  We have
also fixed the quantization concept -- classical canonical
quantization for a theory with a discrete number of steps of freedom.
Momentum quantization is part of this concept.  Moreover, the number
of ether atoms will not vary in this scheme too.  Thus, we fix an
extremely simple, very special class of underlying microscopic
theories.
\footnote{It may be argued that we are unable to do experiments in this
domain, therefore, it is also unreasonable to make a hypothesis about
nature in this domain.  All we can do is to describe their
universality class.  Nonetheless, even if we are unable to do
experiments, we can use general principles as Ockham's razor to make a
choice.  Moreover, there is no need to make a definite choice.  A
situation where we have several metaphysically completely different
models which are all in ideal agreement with experiment would be
satisfactory too -- it gives a clear account about the boundaries of
scientific research.}

Another property of atomic ether theory is especially interesting: the
interpretation of the ``ether density'' $\rho=g^{00}\sqrt{-g}$ as the
number of ``ether atoms'' per volume.  This leads to an interesting
prediction for the cutoff:

\[ \rho(x) V_{cutoff} = 1. \]

The point is that this prediction is different from the usual Planck
length $a_P$.  This can be illustrated with the example of the
``homogeneous universe'' solution (see \S{ } \ref{Friedman}).  In this
solution, $\rho$ remains constant in time.  Therefore, the cutoff
length remains constant in time too.  On the other hand, our rulers
shrink.  Thus, the cutoff length is not constant in our cm scale, it
cannot be the Planck length.  From point of view of our rulers, the
cutoff length is expanding.  For $\Upsilon<0$ or small enough
$\Upsilon$ the cutoff was below Planck scale in the past, for
$\Lambda>0$ or small enough $\Lambda$ it will be greater than Planck
scale, but even greater than the cm scale, in future.  Thus, in this
case we will be able to observe in a far away cosmological future the
effects of the atomic ether.

\section{Canonical atomic ether quantization}

Ether theory suggests to solve the ultraviolet problems with explicit,
physical regularization, based on the idea of an atomic ether.  This
idea leads to the following {\em canonical atomic ether quantization
scheme\/}:

\begin{enumerate}

\item First, we need the full continuous ether theory, that means, a
continuous ``theory of everything'' (TOE) which describes the complete
ether.  This in no way implies that we need some ``grand unification''
to obtain this TOE.  There is no requirement that there should be only
one unified force.  We can as well try to start with GET + SM.

\item Then, we have to find an atomic ether model which gives the TOE
in the large scale limit.  There are obviously many of them, because
the large scale limit fixes only the universality class.  This
uncertainty remains as long as there is no experimental evidence in the
short distance domain.  Therefore simple standard models would be
sufficient.

The possibly critical problem is to prove that the large scale limit
of the discrete model is indeed the original theory.

\item Then, this classical atomic ether theory has to be quantized
in the canonical way.  If the atomic model is close enough to usual
atomic models of usual condensed matter, this step will be
unproblematic.  Note that in this case we can apply also Bohmian
mechanics.  Once this step has been finished, we have quantized
gravity.

\item  Now, for actual quantum computations it is necessary to
derive the large scale limit of the quantum atomic ether theory, which
will be quantum field theory.

Note that this step is already purely phenomenological and not
necessary for the theory itself.  That mean, problems which appear in
this last step are problems of large scale approximations in classical
quantum theory, but not fundamental problems of quantum gravity.

\end{enumerate}

Now, for this quantization program we have to distinguish different
problems.  The first problem is if this scheme works at all.  Here, we
are in a very good situation.  We have an example in reality where the
the whole quantization scheme is realized -- superfluid $^3He-A$.  In
the theory of $^3He-A$, we obtain in the large scale limit the most
important ingredients of the current standard model.  Volovik writes:
``In this sense the superfluid phases of $^3He$, especially $^3He-A$,
are of most importance: the low-energy degrees of freedom in $^3He-A$
do really consist of chiral fermions, gauge fields and gravity''
\cite{Volovik}.  Thus, if we do not have an (uncommon among
physicists) desire for mathematical rigor, for at least one field
theory with gravity, fermions and gauge fields an atomic ether
quantization scheme works.  Moreover, it works in reality.

This is very important -- the existence of such a model in reality
gives certainty that the program may be realized, and it provides
suggestions how this has to be done.

\subsection{Regularization using a moving grid}\label{moving_grid}

Let's see how this works on the example of our second step -- the
derivation of an atomic model for a given continuous ether theory.

Without the ether interpretation, it would be natural to try to
regularize the theory with a regular lattice, following lattice gauge
theory.  Based on the space-time interpretation, we would try to
develop a discrete variant of geometry -- something like the Regge
calculus or dynamical triangulations \cite{Loll}.

The ether interpretation suggests something different -- a
discretization which remains as close as possible to the real atomic
grid.  That means, we do not have to use a regular, static lattice.
Instead, we have to use a grid with the following properties:

\begin{itemize}

\item The grid node density is the ``ether density'' $\rho$:

\[ \sum_{x_k\in V} 1 \approx \int_V \rho d^3x \]

\item The grid moves, with ``ether velocity'' $v^i(x_k)$;

\end{itemize}

Without additional information about the other material properties of
the ether we cannot say anything reasonable about their
discretization.  But in the case of the Minkowski space-time, this
prescription reduces to a homogeneous, static grid.  Therefore, we can
use the lectures of lattice gauge theory \cite{Gupta} to understand
how to discretize them.

Let's consider shortly some details: as a first step to obtain an
atomic model we have to switch from Euler (local) coordinates to
Lagrange (material) coordinates.  Once GET is defined by a Lagrange
formalism, the first interesting problem is if this transformation is
possible in the Lagrange or Hamilton formalism too.  This is known to
be possible in hydrodynamics.  In the Hamilton formalism this can be
done with a canonical transformation \cite{Broer}, \cite{Saarloos},
\cite{Griffa}.  In the following we assume that this is
possible in GET too.

The next important step is the discretization.  Here, it is useful to
have in mind that we want to quantize the theory later.  Therefore, to
be able to use canonical quantization, we need a Lagrange or Hamilton
formalism for the discrete theory.  For this purpose, it is not
reasonable to discretize the equations them-self.  Instead, it is much
more reasonable to discretize the Lagrange function and to define the
discrete equations as Euler-Lagrange equations for the discrete
Lagrangian.

The usual method to obtain a discrete function on a grid is the finite
element method.  In this method, we define functions on the grid as a
subspace of the space of all functions.  In the simplest case, this is
the space of piecewise linear function on the simplices.  These
functions are uniquely defined by their function values on the grid
nodes: $f(x_k) = f_k$.  This defines an embedding of the grid
functions into the space of all functions.  In the other direction, we
can use orthogonal projection of this subspace to define the discrete
image of a continuous function.  Thus, the function values in the
nodes $f_k$ are not defined by the function values of the original
continuous functions, but by integral formulas: $f_k = \int
f(x)\chi_k(x)dx$, where $\chi_k(x)$ is the piecewise linear function
defined by $\chi_k(x_l) = \delta_{kl}$.

In our case, we have an interesting modification of this method: the
density $\rho(x)$ should not be described by a variable grid function
$\rho_i$.  Instead, an integral containing $\rho$ should be
interpolated on the grid in another way.  The simplest way would be

\[ \int_A f(x)\rho(x) dx \to \sum_{x_k\in A} f_k \]

Therefore, in the discrete Lagrange formalism the density simply
disappears.  Now, there are various variants of this method, and which
is the best one depends on the material properties of the ether.
Therefore, further specification of the scheme in this general context
does not seem to be justified and has to be left to future research.

\subsection{Constraints and conservation laws in a moving grid}

Now, it is reasonable to ask about the advantage of this type of
discretization, for example in comparison with a regular, static
lattice.  As far, the only argument was that this looks more natural
from point of view of the atomic ether interpretation.

But, if we look at the remaining problems, we observe that they become
essentially simplified.  Indeed, once we have a well-defined discrete
Lagrange formalism, the most serious remaining quantization problem
are constraints.  Unfortunately, without specification of the material
properties we cannot say anything about possible constraints related
with these material properties.  But there are well-known constraints
in GR, and as we have seen in \S{ } \ref{constraints}, they remain to
be constraints in GET.  Moreover, the conservation laws them-self are
constraints too.

Now, these constraints essentially change their character.  The
continuity equation simply disappears.  It is no longer an equation of
the discrete theory.  The density $\rho$ is no longer part of the
equations.  Instead, the continuity equation becomes a tautology --
the number of grid nodes remains constant. Moreover, the Euler
equations become second order equations: the first order derivative of
$g^{0i}\sqrt{-g}$ becomes a second order derivative of the grid node
position:

\[\partial_t(g^{0i}\sqrt{-g})=\partial_t(\rho v^i) \to m\ddot{x^i} \]

Moreover, we have seen in \S{ } \ref{constraints} that the other
constraints also may be explained by the requirement that there are no
second order equations for $\rho$ and $\rho v^i$.

Therefore, the constraint problem essentially simplifies.  Closely
related with the constraints is the question how the conservation laws
are realized.  The conservation of ether particles in this approach is
realized automatically, as the conservation of the number of grid
nodes.  Therefore, we have no ``quantum fluctuations'' of the ether
particle number.

Note that this property holds in the fundamental, atomic ether theory.
We do not make any claim about the large scale quantum field theory
approximation and the behavior of a field operator $\hat{\rho}(x)$
which fulfills an operator version of the continuity equation.  This
approximation is nothing we have to care as long as we consider
fundamental questions.

\subsection{Universality}

Now, considering the previously discussed program, it seems to suggest
that almost every ether theory may be quantized in this way.  But a
completely different question is how a typical ether theory looks
like.  For this question, the consideration of long distance
universality is important.

The first point of long distance universality is that very different
atomic theories can have the same large scale limit.  ``Long distance
universality is a well-known phenomenon from condensed matter physics,
where we know that a ferromagnet, a liquid-gas system and a
super-conductor may exhibit identical long range properties (phase
diagram, critical exponents, etc.)''  \cite{Jegerlehner}.

This point is important for the justification of the ``moving grid''
method -- it suggests that it is not very meaningful to search for the
``true'' atomic theory without experimental evidence in this domain.
All we can do is to search for a theory which is sufficiently simple
and natural in comparison with their competitors.  Even without
experimental evidence Ockham's razor may be used to choose between
theories.

The other side of large scale universality is that the theories which
appear as large scale approximations have some very typical
properties.  Existing research in this domain has already given
important and interesting results: ``The extraction of the leading low
energy asymptote is equivalent to the requirement of renormalizability
of S-matrix elements, and this has been shown to be necessarily be a
non-Abelian gauge theory which must have undergone a Higgs mechanism
if the gauge bosons are not strictly massless. ... only a
renormalizable field theory can survive as a tail, the possible
renormalizable theories on the other hand are known and are easy to
classify'' \cite{Jegerlehner}.

At a first look, this seems to be in conflict with our atomic ether
quantization scheme.  There seems to be no point where similar
restrictions appear.  But there is no contradiction -- these are
simply different questions.  One question is if a quantization in this
way is possible in principle.  Another question is if a theory has the
{\em typical\/} properties of a large scale limit of an atomic ether
theory.  The answer for the first question may be very well a positive
one, but the related atomic ether theory may require a very strange
conspiracy of their coefficients.

But, if we have seen, for the basic ingredients of the SM -- fermions
and gauge fields -- the situation is very nice.  It is well understood
in theory why such renormalizable theories appear as large scale
limits of a typical atomic ether theory, and we have observed them in
reality in condensed matter, in $^3He-A$.

Considering all these facts, it seems likely that this scheme works,
and that problems which appear on this way may be solved.  Instead,
the problem of canonical GR quantization we discuss in \S{ }
\ref{superposition} suggests that a quantization without a fixed
background fails in principle.

\section{Comparison with canonical quantization of general relativity}

It is interesting to compare our canonical program for GET
quantization with the real way of development of quantization programs
for general relativity, especially the canonical program.  The point
is that the progress of the canonical quantization program is much
more in agreement with ether philosophy than with general relativistic
philosophy.  Let's consider the different steps in the standard
canonical quantization approach:

\subsection{ADM formalism}

The ADM decomposition \cite{ADM} is essentially the decomposition of
$g_{\mu\nu}$ into $\rho, v^i, p^{ij}$.  Therefore, it is an essential
part of the GET approach.

This decomposition is in obvious disagreement with relativistic
philosophy.  Space and time are no longer considered as a unit, they
are separated.  We have a special time coordinate t(x).  Moreover, the
general spacetime manifold becomes subdivided into a product $S
\otimes {\Bbb R}$, that means, changes of topology in time are
excluded.  This modification of relativistic metaphysics is so
important that the ADM formalism is often considered as a different
interpretation of general relativity -- geometrodynamics.  In this
interpretation, GR no longer describes a spacetime, but the evolution
of three-dimensional geometries.

Of course, the Hamilton formalism of general relativity and that of
ether theory remain quite different.  In GET, the Hamiltonian is not a
constraint.  We have equations for the preferred coordinate $T(x)$
used to define the foliation.  And we have additional physical steps
of freedom: density and velocity of the ether.

\subsection{Tetrad and triad formalism}\label{triad}

The next important step is the introduction of the tetrad formalism.
In this formalism, the metric $g^{\mu\nu}$ becomes derived.  We have a
tetrad field -- four vector fields $e^\mu_a(x)$ which form an
orthonormal basis in each point, so that

\[g^{\mu\nu}  = e^\mu_a e^\nu_b \eta^{ab}. \]

This is necessary for the incorporation of fermions into general
relativity.  Obviously this step is also a gross violation of
relativistic ideology.  Indeed, according to this ideology the field
$g^{\mu\nu}$ has already a fundamental interpretation, it defines
spacetime.  The other argument in favor of this procedure is that GR
becomes a gauge theory, with the gauge group SO(3,1).

The next step on this way is the combination of above approaches: The
time coordinate of the ADM composition is used to fix the time-like
tetrad vector.  After this ``time gauge'' we have a triad -- three
vector fields in space, with compact gauge group SO(3).  The technical
problems with non-compact gauge groups are a main argument for this
choice.

Now, a triad field looks very natural from point of view of ether
theory.  If we imagine the ether as a crystal, the three triad vector
fields may be considered as defining locally the orientation of the
crystal structure.  Thus, to introduce triad variables may be an
interesting possibility for canonical GET quantization.  Of course, as
in the case of the ADM variables, we cannot take the formulas as they
are, because we have a different Hamiltonian and different steps of
freedom.

Instead, to consider this triad formalism as something natural from
point of view of relativistic ideology seems impossible.

\subsection{Ashtekar variables}

The next step is a canonical transformation to Ashtekar variables
\cite{Ashtekar} which simplifies the constraints.
There are two variants of the Ashtekar formalism: the first was a
complex formalism.  In this complex formalism, an additional
simplification of the Hamiltonian constraint happens.  For this
advantage it is necessary to pay with the problem of ``reality
conditions''.  But the problem with these reality conditions was too
hard, that's why following Barbero \cite{Barbero} the real version of
the Ashtekar formalism is preferred now.  This seems to be a good idea
from point of view of the ether approach -- the real variant of the
formalism clearly better fits into ether ideology in comparison with a
complex formalism.

The physical meaning of the Ashtekar variables is not obvious at all
in the relativistic approach.  On the other hand, we already know that
we have to do something similar, with similar results, but with clear
physical interpretation in canonical ether quantization: the
transformation from Euler to Lagrange coordinates.  As we have already
mentioned, this is a canonical transformation, and it results in a
simplification of the constraints.

\subsection{Discrete models of geometry}

An important part of existing attempts to quantize gravity are
discrete models.  In some sense, discrete models are also not in ideal
fit with classical spacetime ideology.  It would be much more natural
to have a continuous spacetime for all distances.  But the problems
with non-renormalizability suggest to use discrete regularizations.

On the other hand, the consideration of discrete models is a natural
part of canonical ether quantization, with clear physical motivation:
an atomic ether theory.  Of course, the grids used in ether theory are
three-dimensional grids in a standard Newtonian space, moving in
continuous time.  The position of the grid nodes are steps of freedom
of the ether.  These steps of freedom do not exist in the purely
geometrical approaches.  Nonetheless, we can learn from these
approaches how to discretize the geometric steps of freedom.

For example, it seems quite natural to use the basic ideas of Regge
calculus \cite{Regge} to discretize the pressure $p^{ij}$.  We obtain
a discretization where the pressure $p^{ij}$ is described by a scalar
on each edge between neighbor nodes.  This discretization has a
natural interpretation as the force between neighbor atoms.

\subsection{Summary}

The interesting observation is that, while searching for a way to
quantize general relativity, most success has been reached in a
direction which is in no way close to standard GR ideology:

\begin{itemize}

\item introduction of a preferred time and a Hamilton formalism;

\item introduction of other variables so that the metric
$g_{\mu\nu}$ no longer fundamental;

\item canonical transformations to simplify the constraints;

\item discretization of the theory;

\end{itemize}

Instead, all these steps are quite natural in the ether approach.  We
have formulated most of them in our canonical ether quantization
program.  The steps of freedom in GET are different from the steps of
freedom in canonical GR, and the formulas for the GET approach have yet
to be worked out.  Nonetheless, our observation suggests that the
canonical ether quantization concept is on the right way.

\section{Quantum field theory} \label{QFT}

Now, the question how to quantize an ether theory is conceptually
completely different from the quantization of GR.  The main question
we want to consider here is if this leads to differences in
semi-classical QFT.

In principle, the way we have to quantize continuous ether theory is
to quantize a discrete atomic ether model in a canonical way and then
to consider the large scale limit.  Thus, we have to quantize gravity
similar to the quantization of hydrodynamics by extrapolation of the
microscopic theory, as done by Landau \cite{Landau}.  But it has been
found (Davydov \cite{Davydov}) that the same result may be obtained by
canonical quantization, without using microscopic theory.  Therefore,
without having reasonable microscopic models, it is reasonable to
apply canonical quantization to the continuous GET equations.  To
consider microscopic models seems necessary only for a better
understanding of the way we have to regularize the infinities.  For
example, we can learn why the renormalization of the vacuum energy is
justified.  This can be seen using superfluid $^3He$ as a model
(Volovik \cite{Volovik}).

Once we use canonical quantization, it is no wonder that we obtain the
same formulas as usual in quantum field theory.  Nonetheless, some
remarks seem to be interesting.

First, even if in our covariant formulation the preferred coordinates
formally appear as fields $X^\mu(x)$, this does not mean that they
should be quantized as scalar fields.  This would be a serious
misunderstanding about the purpose of the covariant formulation.
Instead, the $X^\mu$ remain classical preferred coordinates.  This is
an immediate consequence of the basic idea for quantization: to
quantize a microscopic atomic model in a canonical way, using
classical Schr\"odinger theory.

\subsection{Semi-classical quantization of a scalar field}

Let's consider as an example the canonical quantization of a scalar
field on a classical GET background.  We have the Lagrangian

\[\label{defLscal}
{\cal L} =
{1\over2}\sqrt{-g}(g^{\mu\nu}\phi_{,\mu}\phi_{,\nu}-m^2\phi^2) \]

We have a well-defined preferred frame defined by the coordinates
$X^\mu$, and we quantize the field in this frame.  Note that canonical
quantization is a very artificial procedure from point of view of
general relativity - it destroys its covariance ideology.  Instead, it
is a very natural procedure from point of view of ether theory.  Using
the standard formalism of canonical quantization, we obtain

\[\label{defpiscal}
\pi = {\partial{\cal L}\over\partial\phi_{,0}}
= \hat{g}^{0\mu}\phi_{,\mu} \]

\[\label{defHscal}
{\cal H}=\pi\phi_{,0}-{\cal L}
={1\over2} (\hat{g}^{00})^{-1} (\pi-\hat{g}^{0i}\phi_{,i})^2
-{1\over2}\hat{g}^{ij}\phi_{,i}\phi_{,j}
+ {m^2\over2}\phi^2\sqrt{-g}
\]

Note that these expressions look beautiful in the original ether
variables too:

\[\pi = \rho \phi_{,0} + \rho v^i \phi_{,i} \]

\[
{\cal H}={1\over2}(\rho^{-1}\pi^2 - 2 \pi v^i\phi_{,i} -
p^{ij}\phi_{,i}\phi_{,j}) + {m^2\over2}\phi^2\sqrt{-\rho|p^{ij}|}
\]

As we see, our ADM-like decomposition is in good agreement with the
canonical formalism.  We define now $\phi$ and $\pi$ as operators with
the standard commutation rules ($\hbar=1$):

\[\label{defcancomm}
  [\phi(x), \pi(y) ] = i\delta(x-y)
\]

As we see, this definition does not depend on the gravitational field.
This is an important observation.  The background space, its affine
symmetry, the related Hilbert space for the field $\phi(x)$, the
commutation relations and the algebra of observables on this Hilbert
space do not depend on the gravitational field.  This does not seem to
be important in semi-classical theory, but it becomes very important if
we consider superpositions of gravitational fields (see appendix
\ref{superposition}).  In this case, the definition of the Hilbert
space may be used as it is, and scalar products between states defined
for different gravitational fields are well-defined.

This is a very important difference between quantization of GR and
GET.  In GR, the spacetime points and therefore the Hilbert spaces for
the fields $\varphi(x)$ have no independent meaning.

\subsection{Particle operators and vacuum state}

One of the main lectures of quantum field theory is that the
fundamental object are the fields, not the particles.  The notion of
particles is derived, secondary. \footnote{``In its mature form, the idea of
quantum field theory is that quantum fields are the basis ingredients
of the universe, and particles are just bundles of energy and momentum
of the fields'' \cite{Weinberg}.  What ``quantum field theory uniquely
explains is {\em the existence of different, yet indistinguishable,
copies of elementary particles.\/} Two electrons anywhere in the
Universe, whatever their origin and history, are observed to have
exactly the same properties.  We understand this as a consequence of
the fact that both are excitations of the same underlying ur-stuff,
the electron field.  The electron field is thus the primary reality''
\cite{Wilczek}.}
GET does not question this insight.  Instead, in the canonical GET
quantization scheme this becomes exceptionally obvious.  The classical
continuous ether is described by continuous fields -- properties of the
ether.  The fields are fundamental.  Their description does not depend
on the gravitational field -- as we have seen, the Hilbert space for
the quantum field $\varphi(x)$ is defined independent of the
gravitational field.  On the other hand, the notion of particles and
the vacuum state do not appear in a gravity-independent way.  For the
vacuum state we have a natural definition: it is the state with
minimal energy.  But the Hamilton operator depends on the
gravitational field, therefore, the definition of the vacuum state and
the notion of particles too.  In the case of a constant metric
$g^{\mu\nu}$ particle operators are defined by the formulas:

\begin{eqnarray*}
\phi_k  &=& \int e^{ikx}\phi(x)dx\\
\pi_k   &=& \int e^{ikx}\pi(x)dx \\
a^{+}_k &=& {1\over\sqrt{2\omega_k}}(\pi_k -i(\hat{g}^{0i} k_i
- \omega_k)\phi_k),\\
a_k     &=& {1\over\sqrt{2\omega_k}}(\pi_k -i(\hat{g}^{0i} k_i
+ \omega_k)\phi_k),\\
\omega_k^2 &=& \hat{g}^{00} (-\hat{g}^{ij}k_ik_j+m^2\sqrt{-g})
\end{eqnarray*}

with

\begin{eqnarray*}
H = \int {\cal H} dx &=& {1\over2} \int \pi_k^2 + \omega_k^2 \phi_k^2dk\\
\left[a_k,H\right] &=& \omega_k a_k \\
a_k |0\rangle &=& 0
\end{eqnarray*}

Now, in the case of a non-trivial gravitational field, these particle
states are no longer eigenstates of the the Hamilton operator. They
interact with the gravitational field.  Nonetheless, they remain to be
an approximation.  We can introduce wave packets:

\begin{eqnarray*}
\phi_{kx} &=& \int e^{iky-\sigma(y-x)^2}\phi(y)dy\\
\pi_{kx}  &=& \int e^{iky-\sigma(y-x)^2}\pi (y)dy\\
a^{+}_{kx}&=&{1\over\sqrt{2\omega_{kx}}}
     	(\pi_{kx}-i(\hat{g}^{0i} k_i-\omega_{kx})\phi_{kx})\\
a_{kx}    &=&{1\over\sqrt{2\omega_{kx}}}
     	(\pi_{kx}-i(\hat{g}^{0i} k_i+\omega_{kx})\phi_{kx})\\
\omega_{kx}^2 &=& \hat{g}^{00} (-\hat{g}^{ij}k_ik_j+m^2\sqrt{-g})
\end{eqnarray*}

We obtain

\begin{eqnarray*}
[a_{kx},H]     	 &\approx& \omega_{kx} a_{kx} \\
a_{kx} |0\rangle &\approx& 0
\end{eqnarray*}

Here the vacuum state $|0\rangle$ remains to be defined as the state
with minimal energy, it's expression using the local particle
operators becomes an approximation.

This local definition of particle is useful for comparison with
existing semi-classical field theory (cf. \cite{Birrell}).  In this
theory, we have the problem how to define the vacuum state and the
Fock space.  It is usually solved by definition of a set of observers.
For these observers, the vacuum state is the state where they do not
observe particles.

Now, a similar problem does not appear in our canonical scheme.  for
our scalar field we have a natural choice -- the vacuum state as the
state with minimal energy.  But this choice may be understood in a
similar way as the definition of a set of preferred observers -- the
observers which are in rest compared with the preferred frame.  Now,
our local particle operators may be interpreted as the particle
operators which are important for the local observers and their
particle detectors.  In the vacuum state they do not observe
particles.  Therefore, our definition of the vacuum is in agreement
with the definition of the vacuum state related with the set of
preferred observers which are in rest.

\subsection{Different representations}

The major technical problem in quantum field theory on a curved
background $(M,g_{\mu\nu})$ is the existence of infinitely many unitarily
inequivalent representations of the canonical commutation relations.
Isham \cite{Isham} describes it in the following way: ``The real
problems arise if one is presented with a generic metric $g_{\mu\nu}$, in
which case it is not at all clear how to proceed. A minimum
requirement is that the Hamiltonians $H(t)$, or the Hamiltonian
densities should be well-defined. However, there is an unpleasant
possibility that the representations could be $t$ dependent, and in
such a way that those corresponding to different values of $t$ are
unitarily inequivalent, in which case the dynamical equations are not
meaningful.''

Now, GET gives all what may be wanted to prove that this does not
happen: we have a simple equation for the metric (the harmonic
equation), conservation of some important quantities (ether mass and
momentum), we can use inequalities for $\rho$ and p of type
$\varepsilon<\rho<R$ (which may be interpreted as boundaries for the
validity of GET) if necessary.  Possibly this will be sufficient to
solve this technical problem.

On the other hand, this problem appears also in thermodynamics for
states with different temperature -- something which in reality
sometimes changes in time.  Therefore it would not be strange if the
problem nonetheless remains.  I consider it to be an artifact of the
limit $l_{cutoff}\to 0$.

\subsection{Gauge field quantization}

Let's consider now questions related with the quantization of gauge
fields, at first the simplest case of QED in flat space.  There are
different well-known quantization schemes which may be used to
incorporate the gauge condition.  In the variant of Bjorken and Drell
\cite{Bjorken} the gauge condition (Coulomb gauge) is incorporated
into the configuration space:

\[[\dot{A_i}({\bf x},t),A_j({\bf x'},t)]=-i\delta^{tr}_{ij}({\bf x-x'})\]

The other possibility is to consider a large configuration space

\[[\dot{A_\mu}({\bf x},t),A_\nu({\bf x'},t)]
	=-i\delta_{\mu\nu}\delta({\bf x-x'})\]

and to incorporate the gauge condition as an additional restriction
for the states:

\[ (\partial_\mu A_\mu(x))_+ |\Phi\rangle = 0 \]

This scheme leads to a problem with the interpretation of the particle
operators.  We have

\[ [c_{k\mu} c^\dagger_{k'\nu}] = \eta_{\mu\nu}\delta{k-k'} \]

therefore the role of $c_{k0}$ is reversed: $c_{k0}$ behaves like
$c^\dagger_{kj}$.  Now, there are two variants of the interpretation
of these commutation relations.  In the first, classical, Fermi-Dirac
quantization \cite{Fermi}, \cite{Dirac} we accept that the Lorentz
symmetry is broken in the large space: the vacuum is defined by

\[ c^\dagger_{k0} |\Phi_0\rangle = c_{ki} |\Phi_0\rangle  = 0 \]

In the other, explicitly relativistic variant introduced by Gupta and
Bleuler \cite{Gupta}, \cite{Bleuler} we define the vacuum in the
invariant way

\[ c_{k\mu} |\Phi_0\rangle  = 0 \]

and obtain an indefinite Hilbert space:

\[ \langle \Phi_0| c_{k0} c^\dagger_{k0}|\Phi_0\rangle < 0 \]

Again, we have a conflict between relativistic symmetry and a
fundamental physical principle -- the definiteness of the Hilbert
space.  Of course, it is well-known that these differences do not lead
to observable differences.  Nonetheless, this particular quantization
problem is further illustration of the general picture we have found
in Bohmian mechanics as well as for the local energy and momentum of
the gravitational field: every more fundamental description requires
to break relativistic symmetry.  It is obvious that in this case we
prefer the definite Hilbert space.  Therefore, GET suggests to reject
the Gupta-Bleuler approach and to use, instead, the older,
non-covariant Fermi-Dirac quantization scheme.

The choice between Fermi-Dirac quantization and the scheme used by
Bjorken and Drell is more complicate.  We prefer the Fermi-Dirac
quantization scheme because it is based on the Lorenz condition

\[ \partial_\mu A^\mu = 0. \]

This condition is interesting for GET because of the known analogy
between gauge theory and gravity.  Once GET modifies the understanding
of gravity and the EEP, a similar modification of gauge symmetry would
be natural.  In this scenario, the Lorenz condition would be the
natural candidate for a physical equation, similar to the harmonic
condition in GET.  It also allows a physical interpretation as a
conservation law of some ether property.

The incorporation of exact conservation laws into a field theory is a
subtle thing: while integrals over a finite domain very in time, the
integral over the whole space should be exactly conserved, without any
quantum fluctuations.  An atomic ether model suggests natural ways to
reach this property -- if conservation laws are interpreted as
conservation laws for numbers of atoms, and canonical multi-particle
Schr\"odinger theory is used to quantize the theory, this number is
conserved automatically.  In field theory I don't know such a way.  But
it should be noted here again that if this is a problem, it is a
problem of the field theory approximation and therefore not a
fundamental problem of the ether approach.  Field theory is only an
approximation, their problems are therefore not fundamental problems,
but problems of an inconsistent approximation.

\subsection{Non-abelian gauge field quantization}

We do not consider here the quantization of non-abelian gauge fields.
The reason is not that this seems to be very hard.  It is certainly
not impossible, because they appear in real condensed matter (SU(2) in
superfluid $^3He-A$ \cite{Volovik}).  They may be justified as
renormalizable theories which appear in a natural way in the large
scale limit \cite{Jegerlehner}.  For the development of discrete
atomic ether models we can also use the large amount of experience
with lattice QCD \cite{Gupta98}.  The reason is simply that I have not
considered this domain yet in sufficient detail.

\subsection{Hawking radiation} \label{Hawking}

For an instationary gravitational field the vacuum state and the
particle operators depend on time.  Therefore, the original vacuum
becomes a state with particles.  Once the basic concept remains
unchanged, we obtain the same results:

\begin{theorem}
Let $g_{\mu\nu}(X,T)$ be a background metric in preferred coordinates
$X^i,T$ and let's denote the set of observers in rest compared with
the preferred coordinates the ``preferred observers''.  Then in the
canonical formalism we obtain the same results for Hawking radiation
as in the usual formalism for the preferred set of observers.
\end{theorem}

Indeed, the formalism does not depend on the question if the
background metric is a solution of GET or GR.  The only difference
with the standard formalism is the well-defined choice of the
preferred observers in every moment of time.  But this is simply the
application of the general formalism to this special choice of
preferred observers.

But this does not mean that semi-classical GET predicts Hawking
radiation similar to semi-classical GR.  The equivalence holds only as
long as we use the same metric $g_{\mu\nu}$.  For usual configurations
this can be done, the cosmological constants $\Upsilon$ and $\Xi$ may
be ignored.  But for the interesting case of the gravitational
collapse this is not the case.  We obtain a stable ``frozen star''
without horizon (see \S{}\ref{frozen_stars}).

Therefore, during the collapse we obtain Hawking radiation. But once
the collapse has stopped, the radiation goes away and no new radiation
appears, as for stable stars in GR too.  The remarkable result is that
this does not depend on the actual value of $\Upsilon>0$.  Even for
very small $\Upsilon$ the collapsing star needs only a short time to
reach the critical size of the frozen star.  Once a stable state has
been reached, the radiation disappears.  Stable stars do not radiate.

Therefore, GET predicts no Hawking radiation from frozen stars.  For
small enough ``frozen balls'' this leads to observable differences
between GR and GET.  In GET they remain stable and don't evaporate.

\section{Methodology}

Because of the lack of data, in the domain of quantum gravity
methodological and philosophical questions become much more important
than in other domains of science.  In some sense, they become
decisive.\footnote{Butterfield \& Isham \cite{Butterfield} describe
this situation in the following words: ``there are no phenomena that
can be identified unequivocally as the result of an interplay between
general relativity and quantum theory - a feature that arguably
challenges the right of quantum gravity to be considered as a genuine
branch of science at all! ... theory construction inevitably becomes
much more strongly influenced by broad theoretical considerations,
than in mainstream areas of physics.  More precisely, it tends to be
based on various prima facie views about what the theory should look
like -- these being grounded partly on the philosophical prejudices of
the researcher concerned ... In such circumstances, the goal of a
research programme tends towards the construction of abstract
theoretical schemes which are compatible with some preconceived
conceptual frameworks''.}  Does it mean that it is impossible to find
agreement about methodological issues?  Fortunately, the
methodological concepts proposed by Rovelli \cite{Rovelli} are in good
agreement with the method used here.  The key idea of his methodology
is the following:

\begin{quote} ...  confidence in the insight that came with some
theory, or `taking a theory seriously', lead to major advances that
largely extended the original theory itself.  Of course, far from me
suggesting that there is anything simple, or automatic, in figuring
out where the true insights are and in finding the way of making them
work together.  But what I am saying is that figuring out where the
true insights are and finding the way of making them work together is
the work of fundamental physics.  This work is grounded on the {\em
confidence\/} in the old theories, not on random search of new
ones. ... The `wild' scientist observes that great scientists had the
courage of breaking with old and respected ideas and assumptions, and
explore new and strange hypothesis.  From this observation, the `wild'
scientist concludes that to do great science one has to explore
strange hypotheses, and {\em violate respected ideas\/}.  The wildest
the hypothesis, the best. I think wilderness in physics is sterile.
The greatest revolutionaries in science were extremely, almost
obsessively, conservative.
\end{quote}

Now, we are in full agreement with this ``conservative'' view on
fundamental physics, against the popular ``wilderness''.  Based on
this common methodological background, we disagree mainly in one
point: in the decisions what are the ``true insights'' of the old
theories which should be taken seriously and extended into the domain
of quantum gravity, and which should be explained, derived, and
therefore not extended into quantum gravity.

Now, based on these methodological rules, we present the existence of
a preferred frame as the deep insight of Bohmian mechanics.  In other
parts we criticize the usual ``insights'' of general relativity: the
relativistic notion of ``time'' (\S{ } \ref{time}) and relationalism
(\S{ } \ref{relationalism}).

\subsection{The insights of Bohmian mechanics}

In some sense, the disagreement starts with the definition of the
theories we want to unify in quantum gravity.  Usually this is
presented as the problem of unification of general relativity and
special-relativistic quantum field theory.  But it is completely
ignored that there exists already a quantum theory of gravity --
multi-particle Schr\"odinger theory for the Newtonian interaction
potential.  And, moreover, there is an interesting variant of this
theory -- Bohmian mechanics (BM).  While it is in no way suggested
that the interesting and important insights of quantum field theory
should be ignored, the theories we really have to unify are general
relativity and non-relativistic Bohmian mechanics.

We consider some details of BM in appendix \ref{Bohm}.  This theory
proves that vagueness, subjectivity, and indeterminism of the usual
interpretations of quantum theory are not forced on us by the
experimental facts.  These are very interesting and important insights
into the nature of quantum theory -- in my opinion, insights of much
more fundamental character in comparison with the domain of
applicability of particular space-time symmetries.

The problem with BM is that it requires a preferred frame, in
contradiction with the relativity principle.  In this context, it
seems useful to quote again Rovelli \cite{Rovelli}: ``So, Einstein
{\em believed the two theories, Maxwell and Galileo}.  He assumed that
they would hold far beyond the regime in which they had been tested.
...  If there was contradiction in putting the two together, the
problem was ours: we were surreptitiously sneaking some incorrect
assumption into our deductions.''

Now, is there a possibility to believe above theories?  If we require
Lorentz invariance on the fundamental level, Bohmian mechanics should
be simply rejected and does not give any insight. This certainly
violates the recommendation to take above theories seriously.  The
choice of GET is to preserve Lorentz invariance for observable
effects, but to accept a preferred frame on the fundamental level.

Now, the question is how much this weakens the insights of relativity.
Obviously, we do not destroy the insights of relativity completely.
Instead of ``the stage does not exist'' we obtain ``the stage is not
observable''.  This remains to be an important and non-trivial
insight.

\subsection{Relativity as a theory about observables}

But we want to go farther.  We argue that relativity is only a theory
about observables.  Therefore, nothing changes if we replace ``the
stage does not exist'' with ``the stage is not observable''.

Indeed, if there would be a difference, then there should be a method
to show the existence of unobservable objects.  In classical realism,
there are such methods -- the EPR criterion of reality allows to prove
the existence of such hidden objects.  But this criterion has been
rejected.  Moreover, this is a necessity.  Without the rejection of
the EPR criterion relativity would be simply falsified.  Thus, after
the rejection of the EPR criterion we have no longer a chance to prove
the existence of an unobservable stage.  Lorentz invariance {\em is\/}
only observable Lorentz invariance.  The original principle ``all laws
of nature should be Lorentz-invariant'' has been replaced by ``all
observable effects should be Lorentz-invariant'' at least after
Aspect's experiment.

Therefore, from point of view of relativity the statements ``the stage
does not exist'' and ``the stage is not observable'' are simply
identical.  If we interpret the relativistic insight ``the stage does
not exist'' as ``the stage is not observable'' we do not weaken
relativity.  Instead, we take the relativistic preference for
observables seriously.

Thus, we conclude that the acceptance of a preferred frame is in
natural agreement with the methodology recommended by Rovelli, which
requires to take above theories seriously.

\section{The violation of Bell's inequality} \label{Bell}

Before learning the details of the violation of Bell's inequality, I
have thought that it gives some weak evidence in favour of a preferred
frame.  It was a real surprise for me that the evidence is, instead,
very strong -- its not an exaggeration to name it simply a
falsification of relativity.

The whole problem is a very simple one.  While relativity forbids any
superluminal causal influence, the Lorentz ether allows them, as long
as their observable consequences allow two explanations: ($A\to B$ or
$B\to A$).  In this case, they cannot be used to measure absolute
time, which is forbidden in the Lorentz ether.  Now, Bell's inequality
may be violated only if ($A\to B$ or $B\to A$).  And, because it is
violated, relativity is falsified and we have to return to the Lorentz
ether.  A very nice, interesting but simple example of an indirect
existence proof.

But, instead of accepting this elementary experimental falsification,
the simple but fundamental principles used in this proof are
questioned, even rejected.  The argumentation used in this context is
a classical example of immunization.  Some arguments are nonsensical
enough to describe the situation as a ``flight from reason in
science'' (Goldstein \cite{Goldstein}).  Fortunately, reading Kuhn's
``structure of scientific revolutions'' \cite{Kuhn} has recovered my
optimism about the presence of reason in science.  Kuhn observes that
paradigms are never falsified by experiment alone.  Their rejection
always requires another paradigm for replacement.  And, without GET,
the ether paradigm was simply not a viable competitor.

But with GET the ether paradigm becomes a reasonable competitor of
relativity.  Now, with GET as the background, the violation of Bell's
inequality should be reinterpreted.  From point of view of competition
between ether and relativity it becomes a simple and beautiful
experimental falsification of relativity.

Of course, there seems no necessity to copy the well-known proofs of
the various variants of Bell's inequality.  Nonetheless, to explain
some features (like the simplicity of the theorem itself, the
classical character of the decisions and the observations and the
existence of applications) the simplest way seems to be a simple
``proof for schoolboys'':

\subsection{Bell's inequality for schoolboys}\label{BI-proof}

There are three cards, left, middle, right, with red or black color.
I have to choose them so that:

\begin{itemize}
\item the left and middle cards have the same color;
\item the middle and right cards have the same color;
\item the left and right cards have different color.
\end{itemize}

Obviously, one of the three claims is wrong.  Now, if you open two
cards, you can test one of these claims.  What's the probability p to
detect a wrong claim?  Obviously $p\ge 1/3$.

What if you win only with $p<1/3$?  Obviously, something is
manipulated.  What?  That's easy: after you have chosen the first
card, another card may be manipulated.  For example, imagine one card
is marked.  If you open this card, you hand becomes marked.  And the
other card may be manipulated so that if touched by a marked hand it
changes its color.

Let's try to avoid this possibility for manipulation. We use two rooms
and assume that no information transfer between the two rooms is
possible.  In every room we have three cards, and your team has to
choose one card in every room.  How can you be sure that the cards in
the two rooms are the same?  Very simple, you can ask for the same
card in above rooms as often as you like, and in this case these cards
should always have the same color.  Without information about the
question in the other room your opponents cannot be sure if you ask
about different cards or the same cards.  They have no better strategy
than to use the same predefined color for the same card.

Therefore, we are in the same situation as before, but without the
possibility to manipulate based on the information about the other
question.  Therefore, we have again $p\ge 1/3$.  And, if not, there is
a hidden information transfer from one room to the other.

That's all -- we have proven Bell's theorem.  Let's formulate it in
the following way:

\begin{theorem}[Bell's theorem]
If Bell's inequality $p\ge 1/3$ is violated for measurements at A and
B, then there exists a causal influence ($A\to B$ or $B\to A$).
\end{theorem}

\subsection{The difference between special relativity and Lorentz ether}

As far, we have used only classical common sense, and proven that from
the violation of Bell's inequality follows ($A\to B$ or $B\to A$).
The question is now how relativity and ether theory are involved.

In special relativity, we have no absolute time.  Instead, in
pre-relativistic Lorentz-Poincare ether theory we have an absolute
time, but we cannot measure it.  Because of the ether, moving clocks
are dilated and moving rulers contracted.  Now, the formulas for SR
and Lorentz ether are identical.  Therefore, it is often said that
above theories are identical in their predictions.  But this is not
true.  The violation of Bell's inequality is an interesting
difference.

In special relativity we have Einstein causality.  This is a
consequence of two axioms: that there are no causal loops, and that
causality is a law of nature, and therefore has to be Lorentz
invariant.  Einstein causality is Lorentz-invariant, but a notion of
causality which allows faster than light causal influences leads to
causal loops.  Therefore, any causal influence faster than light is
forbidden.  That means, as $A\to B$, as $B\to A$ is forbidden by
Einstein causality.  Therefore, ($A\to B$ or $B\to A$) is forbidden
too.

But in the Lorentz ether the situation is different.  We have
classical causality.  Therefore, causal influences faster than light
are not forbidden.  There is only one restriction: all observable
effects should be Lorentz-invariant.  Now, it seems to follow that
such faster than light influences cannot have observable consequences.
But that's wrong -- observable effects between space-like separated
events A and B may be very well Lorentz-invariant if they allow two
explanations: ($A\to B$ or $B\to A$).  If in absolute time $t_A<t_B$,
we use the explanation $A\to B$, but if $t_A>t_B$, we use explanation
$B\to A$.  Thus, ($A\to B$ or $B\to A$) is not forbidden in the
Lorentz ether.

We conclude that there is an interesting difference in the predictions
of special relativity and Lorentz ether theory: observable
correlations between space-like separated events A and B which may be
explained by causal influences ($A\to B$ or $B\to A$) are forbidden in
special relativity, but allowed in Lorentz ether theory.

Bell's inequality is a simple example of an effect of this type.  We
have above explanations.  If Alice is able to send information to Bob,
they can always win.  If Bob is able to send information to Bob, they
can always win too.  Thus, if they win with probability $p<1/3$, we
can explain this as ($A\to B$ or $B\to A$).  Violations of Bell's
inequality are obviously observable.  We see that it is not correct to
claim that special relativity and Lorentz ether are equivalent as
physical theories.  They are not.  In special relativity, we can prove
Bell's inequality for space-like separated events.  In the Lorentz
ether violations of Bell's inequality are allowed.

\subsection{Aspect's device}

Now, there is no need to understand how Aspect's device \cite{Aspect}
works.  It's sufficient to consider it as a black box, or, more
accurate, a device consisting of two black boxes, one for each room.
You can press one of three buttons -- left, middle, right -- and it
gives the answer -- red or black.  And if Alice and Bob use this
device, your probability to find the wrong answer is $p=1/4$.  That's
all.

We conclude that we have to reject special relativity and to return to
the Lorentz ether.  Bell's conclusion was similar: ``the cheapest
resolution is something like going back to relativity as it was before
Einstein, when people like Lorentz and Poincare thought that there was
an ether --- a preferred frame of reference --- but that our
measuring instruments were distorted by motion in such a way that we
could no detect motion through the ether.  Now, in that way you can
imagine that there is a preferred frame of reference, and in this
preferred frame of reference things go faster than light.''
\cite{Bell1}

\subsection{Should we question everything?}\label{destruction}

We have presented a very simple proof of Bell's inequality.
Unfortunately proofs are certain only in mathematics.  As far as we
consider reality, nothing is certain.  And, therefore, everything in
this proof may be (and has been) questioned.  This reaction is
justified in a situation where relativity is an unquestioned paradigm
of science.  But it is no longer justified in a situation where we
have a competition between two paradigms -- relativity vs. ether
theory.

In this situation, the previously justified search for explanations
becomes a simple method of destruction -- a method which may be
applied to every experimental falsification of a physical theory: take
an arbitrary sentence in the proof we have presented.  Then, name it
an ``unproven hypothesis'', explain that we ``cannot be sure'' that it
holds, and that ``far away from our everyday experience'' this common
sense assumption is possibly invalid.  That's sufficient, you have
found a ``loophole'' in the proof.  Moreover, you will have a lot of
fun if somebody tries to justify the statement you have questioned --
especially because he cannot be successful.  Whatever he presents, you
have a simple counter-attack: take an arbitrary sentence from his
reply, and ...  you already know.
This is the method described by Rovelli \cite{Rovelli} as
``wilderness'':

\begin{quote}
The `wild' scientist observes that great scientists
had the courage of breaking with old and respected ideas and
assumptions, and explore new and strange hypothesis.  From this
observation, the `wild' scientist concludes that to do great science
one has to explore strange hypotheses, and {\em violate respected
ideas\/}.  The wildest the hypothesis, the best. I think wilderness in
physics is sterile.  The greatest revolutionaries in science were
extremely, almost obsessively, conservative.
\end{quote}

That we have to stop somewhere with the questioning is well-known in
scientific methodology.  Every test of a theory requires to stop
questioning at some point.  That's a well-known fact in scientific
methodology.  For example, Popper \cite{PopperLSD} writes:

\begin{quote}
Every test or theory ... must stop at some basic statement or other
which we {\em decide to accept\/}.  If we do not come to any decision,
... then the test will have led nowhere.... Thus if the test is to
lead us anywhere, nothing remains but to stop at some point or other
and say that we are satisfied, for the time being.
\end{quote}

Note that this is not dangerous at all: we do not have to accept
something forever, in a dogmatical way, we can remove our acceptance
whenever we have reason to doubt.  Now, the question may be only what
are reasonable criteria to stop questioning some principles, at least
for time.  In our case, a simple rule seems to be that an objection
should be justified by something more than the simple and trivial
remark that it is an unproven hypothesis.  Based on this basic rule
let's reject now some well known objections against our resolution of
the violation of Bell's inequality.

Thus, unjustified, pure doubt into some common sense principle,
without reason, that means without independent evidence against the
questioned principle, should be rejected.  Let's look now at the
common objections against the obvious resolution of the violation of
Bell's inequality -- the acceptance of a preferred frame.

\subsection{No application for the violation}

One common counter-argument is that we cannot use the violation of
Bell's inequality for information transfer.  This is really strange,
because to use it for information transfer is forbidden in the Lorentz
ether too.  It is simply in contradiction with the requirement that
two explanations of the violation are possible: ($A\to B$ or $B\to
A$).  If we use this device for information transfer $A\to B$ this is
certainly in contradiction with the explanation $B\to A$.  Therefore,
we have the strange situation that it is considered to be an argument
against a theory that the prediction of this theory holds.

A variant of this argument is that the violation of Bell's inequality
cannot be applied.  That's simply wrong.  In our proof, we have
described an application -- Aspect's device allows Alice and Bob to
improve their coordination and to win in the game with probability
$p=3/4$ instead of $2/3$.  But to improve coordination in similar
game-like situations is the classical application of information
transfer.  Therefore, even if pure communication is impossible -- for
known reasons -- we can apply it in a similar way to improve
coordination.  Moreover, it should be noted that this objection, while
common, is in no way an argument in favour of one of the competitors.

\subsection{Objections against a preferred frame}

One argument which was really justified in the past is that there was
no theory of gravity with a preferred frame comparable with general
relativity.  Now, with the theory presented here this argument is
obsolete.  Of course, all other known arguments against the Lorentz
ether may be presented in this context.  Most of them, again, are
obsolete in GET.

One specific argument is worth to be considered: It is argued that it
is not natural to use two different explanations for the same
observation.  But two Lorentz-equivalent configurations are not ``the
same'' in the Lorentz ether.  That's the general situation in the
Lorentz ether.  In different but Lorentz-equivalent situations all
other observable effects are explained in a different way.  In this
context it is natural to have different explanations for the violation
of Bell's inequality too, much more natural than to have the same
explanation.

Now, you may like or not like the properties of the Lorentz ether --
the question is if ($A\to B$ or $B\to A$) is allowed in the Lorentz
ether or not, if there are arguments against the experimental
falsification of relativity.  To count here metaphysical arguments
against the other competitor is really strange.  If we accept
arguments about metaphysical beauty as decisive against an
experimental falsification, this is simply the end of science.

\subsection{No contradiction with quantum mechanics}

Then where is a whole class of common objections: objections related
with the strangeness of quantum theory.  The strangeness of the double
slit experiment or the quantum eraser seems to suggests that the
preferred frame is not the problem, and that the common sense
principles we have used in our proof are in contradiction with quantum
theory.  Of course, we can argue here that the burden of proof is on
the side of the people who claim that there is such a contradiction.
But we don't have to -- instead, we can prove that there is no such
contradiction.

\begin{theorem}
There is no contradiction between the principles used in the proof of
Bell's theorem and the predictions of non-relativistic quantum theory.
\end{theorem}

The proof is given by Bohmian mechanics -- a hidden variable theory of
quantum mechanics found by David Bohm \cite{Bohm}.  In a ``quantum
equilibrium'', this theory makes the same predictions as
non-relativistic quantum theory, therefore, no prediction of
non-relativistic quantum theory is in contradiction with the
predictions of Bohmian mechanics.  But Bohmian mechanics is in full
agreement with the common sense principles used in the proof of Bell's
theorem.  We consider Bohmian mechanics in appendix \ref{Bohm}.

Thus, ``quantum strangeness'' is not a reason to reject the proof.
All arguments related with double slit experiments, spin, quantum
erasers, the ``wholeness'' of quantum theory and so on are irrelevant.
It gives no support for the thesis that something is wrong with the
proof of Bell's theorem.

\subsection{Objections in conflict with Einstein causality}

There is another class of objections which may be rejected as a whole.
It is based on the assumption that Einstein causality is considered as
a prediction of relativity.  The idea is a detailed consideration of
an assumed faster than light phone line.  Let's assume such a phone
line exists.  In this case, we obviously have to give up relativity as
falsified.  Nonetheless, let's consider the seemingly nonsensical
question how we can prove this based on the observation.  The point is
that we can play here the same game of ``questioning everything''.
Now, in this situation this looks certainly nonsensical.
Surprisingly, we look at the details, we find that we have to use
assumptions which may be and have been questioned in the discussion
about Bell's inequality.

We propose the following criterion: {\em all assumptions and
principles which have to be used to prove that a really working
superluminal phone line falsifies Einstein causality should not be
questioned in the proof of Bell's theorem.\/}

Now, let's look how to prove this.  Let's consider the basic device
which transfers one bit of information.  The FTL phone consists of two
black boxes.  We make some decision and press a button on our box.
Our friend makes a measurement on his side of the box and obtains a
result.  Then, we meet later and compare our decisions with his
observations.  If we observe a significant correlation we conclude
that the phone works.

We see that the experimental situation is very close to the situation
in the violation of Bell's inequality, and really includes parts which
have been questioned:

\begin{itemize}

\item The device is a black box, we don't know how it works.  This is
similar to our presentation of Aspect's device as a black box.
Therefore, every argumentation which requires some insight into this
black box before accepting that some causal information transfer has
happened should be rejected.

\item We essentially use the free will of the experimenter.
Of course, as in Bell's theorem, the experimenter may be replaced by
various other random number generators to obtain certainty that the
input -- the decision of the experimenter -- is not predefined.  But
that's enough.  Every argument which does not accept this as
sufficient to establish the independence of the decision of the
experimenter should be rejected.

\item We can verify the existence of a non-trivial correlation only
after the experiment has finished, using other methods of information
transfer.  It is typical for ``many world'' explanations or
explanations based on ``superpositions of observers'': nothing
non-local happens, only if the observers meet again, with usual
subluminal speed, something collapses and we obtain the correlations
in a miraculous quantum way.  These explanations should be rejected
too.

\item Note also the general objection what we observe only a
correlation.  Having only a correlation, we cannot conclude that there
really exists a causal relation in reality.  We need some principle
which allows to make the step from observable correlations to the
conclusion that there exists a real causal interaction.  Every
criticism which rejects the way this has been done in the proof of
Bell's inequality but does not describe an alternative way to conclude
from observation that a real, causal relation exists should be
rejected.

\item We know that as $A\to B$, as $B\to A$ is forbidden by Einstein
causality.  Therefore, ($A\to B$ or $B\to A$) is forbidden too.  This
is obvious.  But there exists a well-known psychological bias known as
``disjunction fallacy'' \cite{Tversky}: if there are two alternatives,
and above alternatives lead to the same conclusion (in our case: the
falsification of relativity), but we are not sure which alternative
happens, we tend not to make the conclusion.  It seems, this fallacy
is the base for arguments of type ``we are unable to detect the
direction of influence''.

\end{itemize}

Again, our argumentation allows to reject a whole class of common
objections.  Of course, it works only if we want to defend Einstein
causality as a physical prediction of relativity.  It does not work
against the idea of rejection of causality itself, for example
following Price \cite{Price94}.

\subsection{Discussion}

Bell's inequality is a prediction of SR which does not hold in LET.
Therefore the violation of Bell's inequality falsifies SR and requires
to accept a preferred frame.  The proof is indirect but so simple and
straightforward that it is hard to imagine a stronger indirect
falsification of a physical theory.

The consideration of the common counter-arguments has not given any
serious loopholes.  Instead, most of the common arguments are part of
two classes we have rejected: objections based on assumed
contradictions with quantum principles, and objections which, if
accepted, would allow to immunize relativity even if we have working
superluminal phone lines.

What remains is the possibility to question everything.  Certain
proofs exist only in mathematics, not in physics.  Immunization of a
physical theory is always possible.  In principle, objections are
always possible.  Last not least, we are talking about reality, not
about pure mathematics.  But there is no independent evidence which
suggests that some part of this proof should be questioned.
Therefore, the rejection of the proof of Bell's inequality has to be
qualified as an ad hoc immunization of relativity.

Nonetheless, in \S{ } \ref{EPR} we consider the EPR criterion in more
detail.  We present there evidence for the thesis that the principles
used in the proof of Bell's inequality -- principles we denote as
``EPR realism'' -- are of more fundamental character than space-time
symmetries.

\section{Conclusions}

General ether theory proposes a paradigm shift back from relativity to
a classical Newtonian background.  It heals the main problems of the
old Lorentz ether:

\begin{itemize}

 \item relativistic symmetry is explained in a general, simple way;

 \item the ether is generalized to gravity;

 \item the ether is compressible, changes in time;

 \item there is no longer an action-reaction problem;

\end{itemize}

In various parts of fundamental physics we find advantages of the new
approach:

\begin{itemize}

 \item we obtain local energy and momentum density for gravity;

 \item the problem of time of quantum gravity is solved;

 \item singularities in physical important situations disappear;

 \item frozen stars instead of black holes solve the information loss
problem of quantum gravity;

 \item we obtain a reasonable dark matter term;

 \item a big bounce instead of a big bang solves the cosmological
horizon problem without inflation;

 \item the EPR criterion of reality holds;

 \item a generalization of Bohmian mechanics into the relativistic
domain becomes unproblematic;

\end{itemize}

Even in the domain of beauty GET seems to be able to compete with GR:
it seems that some of the most beautiful aspects of the mathematical
apparatus of GR obtain a physical interpretation in the context of
GET: harmonic coordinates, ADM decomposition, triad formalism, Regge
calculus.  Especially the beautiful relation between covariance and
conservation laws has to be mentioned here: it gives in GET two nice
expressions for the conservation laws, but in GR it makes energy
conservation highly problematic.

Reconsidering the metaphysical foundations of relativity, we have
found serious flaws:

\begin{itemize}

 \item A reconsideration of the violation of Bell's inequality
suggests its interpretation as a falsification of relativity; the
rejection of the EPR criterion of reality should be interpreted as an
immunization of relativity: except its contradiction with the
relativity principle there is no independent evidence against the EPR
criterion.

 \item A new quantization problem -- named {\em scalar product
problem\/} -- suggests that a really relativistic covariant quantum
theory fails to describe the non-relativistic limit.

 \item The consideration of the ``insight'' of relativity into the
nature of time suggests that it is based on conflation of different
physical notions -- clock time and ``true'' time.  All they have in
common is that they have been named ``time''.

 \item The consideration of general covariance and relationalism shows
that these are not advantages of relativity, but can always be reached
-- by forgetting valuable information.

\end{itemize}

On the other hand, we have been unable to detect serious difficulties
of the new approach: there is no evidence that the introduction of a
preferred frame is problematic in any part of modern physics.

But, of course, a lot of interesting questions remain open, especially
the physical interpretation of gauge fields and fermions, Lagrange
formalisms for condensed matter compatible with the GET scheme, the
details of the related hidden relativistic symmetry in usual condensed
matter.  The unification of the geometrical methods developed for GR
quantization with the canonical quantization schemes for condensed
matter will be interesting for above domains: it defines how to
regularize in quantum gravity, and it gives geometrical interpretation
in condensed matter theory.

\begin{appendix}

\section{Covariant description for theories with preferred coordinates}

The covariant description we consider here is quite simple: we handle
the preferred coordinates formally as ``scalar fields'' $X^\mu(x)$.
This allows to describe a theory with preferred coordinates in a
covariant way.  An interesting point of this covariant description is
that the Euler-Lagrange equations for the preferred coordinates are
the conservation laws.

This is quite obvious and may be considered as a ``folk theorem''.
Nonetheless, confusion about the physical meaning of ``covariance''
seems quite common, and it does not seem to be widely known.  That's
why it seems reasonable to describe the ``weak covariant'' description
we use in more detail.

\subsection{Making the Lagrangian covariant}
\label{covariant_formalism}

We assume that we live in a Newtonian framework.  That means, there is
an absolute Euclidean space and absolute time.  To describe this
absolute background, we use preferred coordinates $X^i, T=X^0$.  As
usual, Latin indices vary from 1 to 3, Greek indices from 0 to 3.  Of
course, to use other coordinates $x^\mu$ is not forbidden, the
coordinates $X^i,T$ are only preferred -- the laws of nature are
simpler in these coordinates.

Now let's consider how to obtain a covariant description starting from
a non-covariant one:

\begin{theorem}
Let $S=\int L(T^{\cdots}_{\cdots},\partial_\mu T^{\cdots}_{\cdots})$
be a functional which depend on components $T^{\cdots}_{\cdots}$ and
first derivatives $\partial_\mu T^{\cdots}_{\cdots}$ of tensor fields
T.

Then there exists a covariant functional $S_c=\int
L_c(T^{\cdots}_{\cdots},\partial_\mu
T^{\cdots}_{\cdots},X^\mu_{,\nu})$ which depends on the components and
first derivatives of the same tensor fields T and on the first
derivatives $X^\mu_{,\nu}$ of four scalar fields $X^\mu(x)$ so that

\[ S(T) = S_c(T, X^\mu_{,\nu}) \]

for $X^\mu(x) = x^\mu$.
\end{theorem}

Proof: In L we replace the tensor components by expressions using the
following replacement rules for all indices:

\begin{eqnarray*}
T^{\cdots\mu\cdots}_{\cdots} &\to&
     	{\partial X^\mu\over\partial x^\nu}T^{\cdots\nu\cdots}_{\cdots}\\
T_{\cdots\mu\cdots}^{\cdots} &\to&
       	{\partial x^\nu\over\partial X^\mu}T_{\cdots\nu\cdots}^{\cdots}\\
T_{\cdots,\mu}^{\cdots} &\to&
       	{\partial x^\nu\over\partial X^\mu}T_{\cdots,\nu}^{\cdots}\\
T^{\cdots\mu\cdots}_{\cdots,\cdot} &\to&
     	{\partial X^\mu\over\partial x^\nu}T^{\cdots\nu\cdots}_{\cdots,\cdot}\\
T_{\cdots\mu\cdots,\cdot}^{\cdots} &\to&
       	{\partial x^\nu\over\partial X^\mu}T_{\cdots\nu\cdots,\cdot}^{\cdots}
\end{eqnarray*}

The matrix ${\partial x^\nu\over\partial X^\mu}$ is the inverse matrix
of ${\partial X^\mu\over\partial x^\nu}$.  We use this property to
express all occurrences of ${\partial x^\nu\over\partial X^\mu}$ by
these rational functions of $X^\mu_{,\nu}$.  For $X^\mu(x) = x^\mu$
these are obviously identical transformations.  Moreover, each
argument is now a covariant scalar.  Indeed, the indices $\mu$ of the
$X^\mu_{,\nu}$ are not tensor indices, but enumerate the scalar fields
$X^\mu(x)$.  But indices of this type are the only open indices in the
expression.  Therefore, being a function of covariant scalar
expressions, the modified function $L_c$ is a covariant scalar
function, qed.

Covariant functions like $L_c$ which do not depend on the fields
$X^\mu(x)$ we name ``strong covariant'', to distinguish them from
``weak covariant'' functions which depend on the fields $X^\mu(x)$.

Of course, what is not part of this theorem is how to interpret
something defined only in the preferred coordinates as a component of
some tensor field.  This is, essentially, the real problem if we have
to make a theory covariant.  There are usually different
possibilities: a 3D scalar may be a 4D scalar as well as a component
of a 4D vector or tensor field.  But this is a question of the
definition of the theory itself.  We are interested here only in a
special way to obtain a covariant description of a well-defined
theory.

\subsection{Conservation laws in the covariant formalism}
\label{covariant_conservation}

This formalism raises two interesting questions.  First, once we have
a covariant formalism, we obtain the known problems with conservation
laws -- Noether's theorem does not give non-trivial conservation laws.
Second, we have four new Euler-Lagrange equations for $S=\int L$ --
the Euler-Lagrange equations for the preferred coordinates $X^\mu$.
What is their physical meaning?  Now, the answer is simple and
beautiful -- the Euler-Lagrange equations for the preferred
coordinates are the conservation laws.  All we need is to look at the
Euler-Lagrange equations for the preferred coordinates:

\begin{theorem}
If a Lagrangian L does not depend explicitly on the coordinates, then
the Euler-Lagrange equation of the related weak covariant Lagrangian
$L_c$ for the preferred coordinates $X^\mu$ defines conservation laws
for the tensor

\[ T^\nu_\mu = - {\partial L_c\over\partial X^\mu_{,\nu}} \]

If the original Lagrangian L is covariant, then $T^\nu_\mu=0$.
\end{theorem}

If there was no explicit coordinate-dependence in the original
Lagrangian, there is none in $L_c$ too.  Note also that for a
covariant Lagrangian $L$ we have $L\equiv L_c$ and does not depend on
$X^\mu$ and $X^\mu_{,nu}$.  This follows from the construction.  The
theorem now follows immediately from the the Euler-Lagrange equation
for the $X^\mu$:

\[ {\delta S\over\delta X^\mu} = {\partial L\over\partial X^\mu}
 - \partial_\nu{\partial L\over\partial X^\mu_{,\nu}} = 0 \]

The results about the existence of conservation laws are equivalent to
Noether's theorem, but the energy-momentum tensor is not the same.
The relation between preferred coordinates and conservation laws is a
much more direct one in this formalism: de facto one line was
sufficient for the proof.  We have not used the other Euler-Lagrange
equations.  That's why we consider this variant of the conservation
law as the more fundamental one.

We conclude: {\em the Euler-Lagrange equations for the preferred
coordinates are the conservation laws.\/}

\section{Relationalism}\label{relationalism}

The confusion about covariance has historical reasons.  Initially
covariance was believed to be a special, distinguishing property of
GR.  Later it has been recognized that every physical theory allows a
covariant formulation.  For example, Fock \cite{Fock} has given a
covariant formulation for special relativity.  A simple way to do this
is to use the curvature tensor of a metric $g_{ij}$.  The equation
$R^i_{jkl}=0$ defines a flat metric in a covariant way.  Once a flat
background has been defined, all partial derivatives of the equations
in the preferred frame may be replaced by covariant derivatives of
this background metric.  This method has been widely used to present
theories of gravity with predefined geometries, for example
\cite{Lightman}, \cite{Ni}.

Nonetheless, this does not mean that this question is completely clear
now.  The situation is confusing.  On one hand, every theory allows a
covariant formulation, on the other hand, there is a non-trivial
symmetry property -- the property we have named ``strong covariance''
in GET.  Rovelli \cite{Rovelli} describes it using the notions
``active vs. passive diff invariance'':

\begin{quotation}
All this is coded in the active diffeomorphism invariance (diff
invariance) of GR.  Passive diff invariance is a property of a
formulation of a dynamical theory, while active diff invariance is a
property of the dynamical theory itself.  A field theory is formulated
in manner invariant under passive diffs (or change of coordinates), if
we can change the coordinates of the manifold, re-express all the
geometric quantities (dynamical {\em and non-dynamical\/}) in the new
coordinates, and the form of the equations of motion does not change.
A theory is invariant under active diffs, when a smooth displacement
of the dynamical fields ({\em the dynamical fields alone\/}) over the
manifold, sends solutions of the equations of motion into solutions of
the equations of motion.
\end{quotation}

This is in agreement with our understanding. The problematic part is
how to distinguish dynamical and non-dynamical fields.  Rovelli
continues:

\begin{quotation}
Distinguishing a truly dynamical field, namely a field with
independent degrees of freedom, from a non-dynamical field disguised as
dynamical (such as a metric field $g$ with the equations of motion
Riemann[g]=0) might require a detailed analysis (for instance,
Hamiltonian) of the theory.
\end{quotation}

That means, it is assumed that an analysis of the equations of the
theory -- the dynamics -- allows to distinguish the non-dynamical and
the dynamical steps of freedom of the theory.  The example of GET and
its relation with general relativity with four scalar dark matter
fields (GRDM) allows to proof that this is impossible:

{\bf Thesis:} {\em It is impossible to distinguish a non-covariant
theory with a ``preferred frame in disguise'' from a ``truly''
covariant physical theory by evaluation of the equations of motion and
the Lagrange formalism.\/}

Indeed, in GET, the preferred coordinates $X^\mu$ are non-dynamical,
and their presentation as ``dynamical fields'' $X^\mu(x)$ fits exactly
with the description of a ``non-dynamical field disguised as
dynamical''.  As we have seen, the equation for the preferred
coordinates $X^\mu(x)$ is simply the harmonic equation, thus, the
usual general-relativistic equation for scalar fields.  On the other
hand, in GRDM the four scalar fields are dynamical.  But the Lagrange
formalism for GET is exactly the Lagrange formalism for GRDM.

A variant of this argument is the consideration of GET with the four
``preferred coordinates in disguise'' $X^\mu(x)$ and a few additional
free scalar fields $\varphi^m(x)$.  Then, looking at the equations and
the Lagrangian, there is no way to tell what are the truly dynamical
scalar fields $\varphi^m(x)$ and what are the coordinates in disguise
$X^\mu(x)$.  We need additional a-priori information, additional
insight.

Once the equations are identical, even a ``Hamiltonian analysis''
(whatever this means in detail) cannot help.  But there is an
interesting point: the Hamilton formalism is different for GET and
GRDM.  In GRDM we have the typical problems of general-relativistic
theories where the Hamiltonian is a constraint.  Instead, in GET we
have a classical Hamilton formalism, and the Hamiltonian is not a
constraint.  But we cannot derive the Hamilton formalism of GET
without the additional information that the fields $X^\mu(x)$ are the
preferred coordinates.  Without this information we cannot decide
which Hamilton formalism is the appropriate one.

Note that our thesis does not mean that there are no differences
between GET and the ``truly relativistic'' theory GRDM.  The
differences we have considered in \S{ } \ref{GRDM}.  Our thesis
is that we need additional insight -- the insight that the fields
$X^\mu$ are preferred coordinates -- to be able to distinguish
a theory with absolutes from a ``truly relativistic'' theory.

\subsection{What is the true ``insight'' of general relativity?}

This observation seems to destroy the whole concept described by
Rovelli \cite{Rovelli}:

\begin{quote}
One of the thesis of this essay, is that general relativity is the
expression of one of these insights, which will stay with us
``forever''.  The insight is that the physical world does not have a
stage, that localization and motion are relational only, that
diff-invariance (or something physically analogous) is required for
any fundamental description of our world \ldots.

In GR, the objects of which the world is made do not live over a stage
and do not live on spacetime: they live, so to say, over each other's
shoulders.
\end{quote}

But, as we have seen, this is not an insight, but a tautological
reformulation.  If the theory has a stage, we can reformulate it and
name the stage a dynamical field.  Instead of ``absolute motion'' we
talk about ``motion relative to the stage field''.  Let's see how
Rovelli describes Einstein's ``insight'':

\begin{quote}
Of course, nothing prevents us, if we wish to do so, from singling out
the gravitational field as ``the more equal among equals'', and
declaring that location is absolute in GR, because it can be defined
with respect to it.  But this can be done within any relationalism: we
can always single out a set of objects, and declare them as not-moving
by definition. The problem with this attitude is that it fully misses
the great Einsteinian insight: that Newtonian spacetime is just one
field among the others.
\end{quote}

The situation between GET and GRDM is the reverse one.  The fields
$X^\mu(x)$ are not just scalar fields among others.  But this is a
non-trivial insight.  If we forget this insight, and the fields
$X^\mu(x)$ are interpreted just as fields among others, we have lost
important information and interesting predictions.  In this way, by
forgetting interesting information, a relational description is always
possible.  Therefore, the existence of a relational description is
only an insight into the mathematical formalism of physical theories
in general, not an insight of GR.  Instead, in GET the existence of
absolutes -- the preferred coordinates $X^\mu(x)$ -- is a non-trivial
insight which leads to interesting predictions: the fields $X^\mu(x)$
may be used as global coordinates, the field $X^0(x)$ is time-like.

\subsection{What are insights which are ``forever''?}

What would be an important insight is that a theory with a certain
type of absolutes is impossible in nature.  But for this insight
general relativity is not enough.  This can be, by its nature, only an
impossibility proof for certain classes of theories.  Such an insight
may be based on certain observational facts, for example, observations
of a worm-hole.  This would be incompatible with a whole class of
theories with flat background.  Such an observation of a non-trivial
topology, for example a worm-hole, would be really an insight which
remains forever.\footnote{Another example of this type of insight is
the violation of Bell's inequality, which excludes a whole class of
theories -- theories which allow the proof of Bell's inequality for
space-like separated observations.  This can be named also an
``insight of quantum theory''.  But this would be a very sloppy
description: the proof of Bell's is classical, and the violation of
Bell's inequality has been observed in real experiments.}  But we have
not observed such non-trivial topology.

A lecture which can be learned from history of science is that
metaphysical preferences like between relationalism and the existence
of a predefined stage should not be justified based on the current
physical theory.  Rovelli himself has presented the best example of
this type -- Newton's insight, related with the famous rotating
bucket, that there exists an absolute space.  This insight, which was
thought to remain forever too, was rejected by Einstein.
In Einstein's theory, we have relationalism, no absolutes.
But this insight may be as well superseded by another theory.
GET proves that this is possible.

There is another example of an interesting metaphysical question where
we tend to use current physical theory: probability versus
determinism.  Here we have even more switches between probabilistic
chaos and determinism.  Bohmian theory (deterministic), quantum theory
(probabilistic), classical mechanics (deterministic), chaos in the
many-particle situation (probabilistic), classical thermodynamics
(deterministic), and chaos in its large scale predictions.  Therefore,
to base the metaphysical decision between chaos and determinism on
current physical theory is highly speculative.  To guess that this
property remains forever is sufficiently falsified by historical
evidence.

There are important insights of general relativity.  For example that
a special physical entity which was absolute in Newton's theory is not
absolute in reality.  It is the entity which defines inertial forces
and causes the difference between a rotating bucket and a bucket in
rest.  This entity is the gravitational field.  This insight into the
nature of gravity and clock time will stay forever.

Another insight is the Einstein equivalence principle.  Because this
is an exact symmetry claim, we cannot be sure that it remains forever
-- but we can be sure that it survives at least as an approximative
symmetry.

\section{The problem of time} \label{time}

It is well-known that the problem of time may be solved by the
introduction of a preferred foliation as in GET: ``in quantum gravity,
one response to the problem of time is to `blame' it on general
relativity's allowing arbitrary foliations of spacetime; and then to
postulate a preferred frame of spacetime with respect to which quantum
theory should be written. Most general relativists feel this response
is too radical to countenance: they regard foliation-independence as
an undeniable insight of relativity.''  \cite{Butterfield}

\subsection{clock time vs. true time}

To meet this argument, we have to consider the ``insight of
relativity'' about the nature of time and the metaphysical aspects of
the ``problem of time'' in more detail.  In some sense, the problem of
time may be discussed based on Newton's definition \cite{Newton}.
Newton distinguishes two notions of time:

\begin{quote}
... I do not define time, space, place, and motion as being well known
to all.  Only I observe, that the common people conceive these
quantities under no other notions but from the relation they bear to
sensible objects.  And thence arise certain prejudices, for the
removing of which it will be convenient to distinguish them into
absolute and relative, true and apparent, mathematical and common.

{I. Absolute, true and mathematical time, of itself, and from its own
nature, flows equable without relation to anything external, and by
another name is called duration: relative, apparent, and common time,
is some sensible and external (whether accurate or unequable) measure
of duration by means of motion, which is commonly used instead of true
time; such as an hour, a day, a month, a year.}
\end{quote}

This definition is in agreement with GET.  Here, we have an
unobservable harmonic ``true time'' $T(x)$ together with the
``apparent time'' $\tau=\int\sqrt{g_{\mu\nu}dx^\mu dx^\nu}$ measured
with clocks.  These two notions of time may be roughly identified with
``time of quantum theory'' and ``time of relativity'':

Indeed, in quantum theory we have no self-adjoint operator for time
measurement.  Instead, it is closely related with the fundamental
aspects of quantum theory: physical quantities have to be measured at
a given time, the scalar product is conserved under time evolution.
Thus, it ideally fits with Newtons ``true time'' and should be
distinguished from ``apparent time''.

On the other hand, time in relativity is {\em defined\/} to be
apparent time.  General-relativistic time is the time measured by
clocks -- ``measure of duration by means of motion''.  And its most
interesting physical property is that it is unequable.  The fit with
Newton's definition of apparent time is simply ideal.

Therefore, if we deny Newton's insight into the different nature of
true and apparent time, and try to develop quantum gravity without
making this distinction, we are immediately faced with the problem to
unify these two different notions of time.  This is the metaphysical
base of the notorious ``problem of time'' in quantum gravity.

\subsection{About positivistic arguments against true time}

The standard relativistic argumentation against this understanding of
time is the rejection of ``true time'' based on positivistic
arguments.  The typical argument is that a physical notion needs an
operational definition.  If we cannot measure something, then it is
not part of physics.

Now, this argumentation is based on positivistic methodology of
science, which has been rejected by Popper \cite{PopperLSD}.
According to Popper's ``logic of scientific discovery'' theory is
prior to observation.  Therefore, the fundamental objects of a theory
are not based on observation, they do not need any operational
definition.  It is not the observation which decides about the
fundamental notions of the theory.  Instead, observation is always
theory-laden.  In Einstein's famous words, it is the ``theory which
decides what is observable''.  Positivism reverts the relation between
theory and observation.  All what is required in science is that the
theory, as a whole, makes a lot of falsifiable predictions.  An
operational definition of some fundamental notions of the theory is a
useful tool to obtain such predictions, but not more.  Positivism is
simply wrong.  Again, in Einstein's words about Bohr's positivism:
"Perhaps I did use such a philosophy earlier, and even wrote it, but
it is nonsense all the same."

Moreover, in our case we do not have to rely on such methodological
considerations.  Instead, we have a beautiful example of a theory with
unobservable ``true time'' -- classical quantum theory.  We have
already mentioned that there is no self-adjoint operator for time
measurement.  As a consequence, no physical clock can provide a
precise measure of time.  There is always a small probability that a
real clock will run backward with respect to it \cite{Unruh_Wald}.
Nonetheless, the time of quantum theory is not only an important part
of quantum theory.  A lot of people have tried hard to remove time
from quantum theory, without success.

The example of time in quantum theory is not only a powerful
illustration of the failure of positivistic argumentation, but answers
the question about the physical meaning of true time: the simple
answer is ``the same as in classical quantum theory''.  The advantage
of this answer is that we do not need any vague metaphysical
considerations about the nature of ``true time''.

\subsection{Relativity as an insight about true time}

On the other hand, positivistic argumentation against true time is not
the only relativistic argumentation.  It seems, a lot of relativistic
scientists acknowledge very well that there is more behind the notion
of time than simple clock measurement.

Indeed, let's for this argument accept the positivistic argumentation
that there is nothing like ``true time'', and the only physically
meaningful notion of time is clock time.  What, in this case, is the
physical meaning of the following -- rather typical -- relativistic
argumentation \cite{UnruhTime}:

\begin{quote}
One often hears that what General Relativity did was to make time
depend on gravity ... Such a dependence of time on gravity would have
been strange enough for the Newtonian view, but General Relativity is
actually much more radical than that.\footnote{We have to disagree
with this claim that a dependence of apparent time on gravity would
have been strange for the Newtonian view.  Last not least, a classical
clock based on Newtonian theory -- the pendulum clock -- obviously
depends on gravity.} ... Rather the theory states that the phenomena
we usually ascribe to gravity are actually caused by time's flowing
unequable from place to place....  Most people find it very difficult
even to imagine how such a statement could be true. ...  That gravity
could affect time, or rather could affect the rate at which clocks
run, is acceptable, but that gravity is in any sense the same as time
seems naively unimaginable.
\end{quote}

As we see, the point is not that apparent time -- time measurement
with clocks -- is influenced by gravity.  This is only the part which
is already ``acceptable'' to people who have not really understood
relativity.  The point which is considered to be the ``naively
unimaginable'' insight of relativity is the identification with
``time'' -- obviously not ``apparent time'', but ``true time''.  Thus,
the non-existing ghost of true time revives here in its full beauty,
as an important, fundamental insight of relativity.

In my understanding we have here a conflict between two common
relativistic argumentations.  On one hand, the rejection of the notion
of ``true time'' and the reduction of time to clock time measurement,
on the other hand the metaphysical identification of the same ``true
time'' with time measurement, presented as an ``insight of
relativity''.  This contradiction shows that there is not much in
support of this ``insight'' except methodological confusion:

Or we accept that there is nothing behind ``time'' except clock
measurement.  Then no non-trivial insight into the nature of ``true
time'' exists.  The (nonetheless very important) insight of relativity
is about the influence of gravity on clocks.

Or we accept that time is more than clock measurement.  Then there is
no empirical base for the identification of clock time with this other
notion of time.  The identification is not only a purely metaphysical
hypothesis without empirical support.  Moreover, this metaphysical
hypothesis is highly questionable, it seems to be based only on
misunderstanding because the two notions are usually described with
the same word -- time.  Simply two completely different notions of
time which have been distinguished already by Newton have been
mingled.

\subsection{Technical aspects of the problem of time}

The metaphysical choice to identify quantum true time with
relativistic clock time leads to subsequent technical problems.  In
different approaches to quantum gravity the problem of time shows in
different ways.  It is not the purpose of this paper to consider them,
we refer here to \cite{Butterfield}, \cite{Isham}, \cite{Baez} for
further details.  But an example may be informative.  In the canonical
approach which seems to be the closest to our approach, some
``internal coordinates'' have to be chosen.  But this leads to a
``multiple choice problem'' \cite{Isham}:

\begin{quote}
Generically, there is no geometrically natural choice for the internal
spacetime coordinates and, classically, all have an equal
standing. However, this classical cornucopia becomes a real problem at
the quantum level since there is no reason to suppose that the
theories corresponding to different choices of time will agree.

The crucial point is that two different choices of internal
coordinates are related by a canonical transformation and, in this
sense, are classically of equal validity. However, one of the central
properties/problems of the quantization of any non-linear system is
that, because of the well-known Van-Hove phenomenon
\cite{VanHove}, most classical canonical transformations {\em
cannot\/} be represented by unitary operators while, at the same
time, maintaining the irreducibility of the canonical commutation
relations. This means that in quantizing a system it is always
necessary to select some preferred sub-algebra of classical
observables which is to be quantized.
\end{quote}

\section{Quantum gravity requires a common background}
\label{superposition}

In this section, we consider an interesting problem of the
quantization of GR.  This problem has been the starting point for the
author.  It strongly suggest that for successful quantization GR
should be modified, especially that it is necessary to introduce a
common background manifold.

The main difference between GR and other theories including SR is that
GR is a ``one world theory''.  It does not have a possibility to
compare different gravitational fields.  In other theories we have
some common background.  This allows to talk about values at ``the
same point'' for different solutions.  In GR, this is not possible.
Another solution may be defined even on another manifold.

This is not problematic because a possibility to compare different
solutions is not necessary in the classical domain.  As long as we
consider classical gravitational fields, there is only one solution we
can observe, there is only ``one world''.  Therefore, in the classical
domain a ``one world theory'' does not cause problems.  But what about
quantum superposition?  In a superpositional state we can consider
different classical solutions in a single state.  Is a ``one world
theory'' sufficient to describe superpositions?  Or is there more
information hidden in a superpositional state?  Especially, is there
information about relative position between the states which are in
superposition?

This simple question may be the key for the understanding of quantum
theory.  There seem to be only two ways to answer this question: yes
or no.  The relativistic answer is a clear no.  Our answer is a clear
yes.  A nice way to formalize the two possible answers has been
proposed by Anandan \cite{Anandan}: the notions of c-covariance and
q-covariance.  We consider superpositions of quasi-classical
gravitational fields.  Now, we have to look what happens if we apply
diffeomorphisms.  We have different fields, and this raises the
question if we are allowed to use different diffeomorphisms for these
different fields or not.  This leads to two ways to generalize
classical covariance: The first, weak generalization is c-covariance
-- it requires covariance only if we apply the same diffeomorphism.
The second, strong generalization is q-covariance -- it allows
different diffeomorphisms for the different states in superposition.

Now, the relativistic answer is unique -- the only appropriate
generalization is q-covariance.  The reason is that c-covariance
requires an essential modification of classical GR.  Indeed, the
Einstein equations of classical GR defines a pair of solutions only
modulo q-covariance.  The GR Lagrangian is also q-covariant.  Thus, to
define the evolution of c-covariant but not q-covariant objects we
simply do not have appropriate equations in classical GR.  Moreover,
even the notion of ``the same diffeomorphism'' is not appropriately
defined in GR.  Different GR solutions live on different manifolds,
and there is no well-defined notion of ``the same diffeomorphism'' on
another manifold.  This is the solution which has been proposed by
Anandan too.

We have chosen the other way.  We think that the description of a
superposition requires additional, non-q-covariant information.
The purpose of this section is to justify this choice.

Let's start with a simple question -- the superposition of a state
with a shifted version of itself:
$|g_{ij}(x)\rangle+|g_{ij}(x-x_0)\rangle$.  Do you really think this
may be simply the original state, not a non-trivial superposition?
I'm not.  But, of course, this simple argument does not seem to be
sufficient.

Fortunately, we can present a much stronger argumentation.  As we see,
already a very simple scattering experiment on such superpositional
states gives observables which depend on ``relative position''.

\subsection{A non-relativistic quantum gravity observable}

In the consideration of quantum gravity, people usually consider two
basic theories: classical general relativity and relativistic quantum
field theory on a fixed background.  But these are not really the
theories which have to be unified -- the gravitational field is
classical in above theories.  The interesting point of quantum gravity
is, of course, the consideration of superpositions of gravitational
fields.  It is, last not least, superposition which makes quantum
theory different from classical statistics.  And this difference will
be the point of our experiment: what we want to measure is the
transition probability which decides if a superposition has been
destroyed by measurement or not.  The problem with semi-classical QFT
is that it does not give a base for such considerations.

On the other hand, it is not at all difficult to compute such
transition probabilities in a reasonable approximation.  For this
purpose, we can use a well-known simple theory which allows to
consider non-trivial superpositions of gravitational fields.
This theory is simply non-relativistic quantum gravity -- classical
multi-particle Schr\"odinger theory with Newtonian interaction
potential.

There is nothing ill-defined with this theory.  From theoretical point
of view it works as well as multi-particle Schr\"odinger theory with
Coulomb interaction potential.  The fact that we have no data is not
really problematic.  Indeed, the theory unifies classical quantum
principles with classical gravity in an ideal way.  If somebody tends
to doubt simply because there are no data, quantum gravity is a
forbidden area for him.  Thus, to assume the correctness of
Schr\"odinger theory in the non-relativistic limit is not problematic.
But, of course, it is a non-trivial decision: {\em We assume that
classical Schr\"odinger theory is the non-relativistic limit of
quantum gravity.\/}

Now, based on this non-relativistic theory we can consider simple
gravitational scattering.  This gives some insight into features of
superpositions of gravitational fields, even if the field itself is
not quantized.  We simply have to consider superpositional states of
an otherwise neutral particle.  We need the particle only as a source
of gravity.  Thus, let's consider a situation like a double slit
experiment, with some superpositional state
$|\psi^1\rangle+|\psi^2\rangle$ of a source of gravity.  For the
interaction with a test particle $|\varphi\rangle$ only the
gravitational field of the source is important.  Thus, it is possible
to interpret the interaction also as an interaction of the test
particle $|\varphi\rangle$ with a superpositional state of the
gravitational field $|g^1\rangle+|g^2\rangle$.

Now, let's make some simplifying assumptions.  First, let the mass M
of the source particle be much greater than the mass m of the test
particle: $M\gg m$.  In this case, the state of the heavy particle is
much less influenced by the interaction than the state of the test
particle. Let's also assume that the state of the source particle is
highly localized: $\psi^1(x)\approx\delta(x-x_1)$,
$\psi^2\approx\delta(x-x_2)$.  In this case, we can use single particle
theory to compute the result of the interaction. Let's denote with
$\varphi^1, \varphi^2$ the solution of the Schr\"odinger equation for
the source particle located in $x_1$ resp. $x_2$.  Then, for the
two-particle problem with the initial values
$|\psi^1\rangle\otimes|\varphi\rangle$ resp.
$|\psi^2\rangle\otimes|\varphi\rangle$
we obtain approximately a tensor product solution
$|\psi^1\rangle\otimes|\varphi^1\rangle$ resp.
$|\psi^2\rangle\otimes|\varphi^2\rangle$.
For the superpositional state
$(|\psi^1\rangle+|\psi^2\rangle)\otimes|\varphi\rangle$
we obtain the solution by superposition:

\[|\psi^1\rangle|\otimes|\varphi^1\rangle
 +|\psi^2\rangle|\otimes|\varphi^2\rangle.\]

But this is equivalent to

\[(|\psi^1\rangle+|\psi^2\rangle)\otimes
  (|\varphi^1\rangle+|\varphi^2\rangle)
 +(|\psi^1\rangle-|\psi^2\rangle)\otimes
  (|\varphi^1\rangle-|\varphi^2\rangle)\]

Now, we are interested in the transition probability
$|\psi^1\rangle+|\psi^2\rangle\to|\psi^1\rangle-|\psi^2\rangle$.
We obtain

\[ p_{trans} ={1\over2}(1 - Re\langle\varphi^1|\varphi^2\rangle)\]

To understand why we are very interested in this transition
probability let's consider the limiting cases: if
$\langle\varphi^1|\varphi^2\rangle=1$, gravitational interaction is
not important, the position of the source particle does not influence
the state of the test particle.  Thus, the superpositional state
remains unchanged, the interaction with the test particle was not a
measurement of position of the source particle.  We have a tensor
product state.  Therefore, we can ignore the test particle and obtain
a pure one-particle state for the source.

In the other limiting case, the resulting states are orthogonal,
$\langle\varphi^1|\varphi^2\rangle=0$, therefore, the transition
probability is $1\over 2$.  We do not have a product state.  If we
ignore the test particle, we do not obtain a pure source particle
state.  Instead, we obtain a classical mixed state:

\[ {1\over 2}(|\psi^1\rangle\langle\psi^1|+|\psi^2\rangle\langle\psi^2|) \]

Thus, the transition probability defines if the interaction was a
measurement which has destroyed the superposition or not.  If there is
something which allows to distinguish a superposition from a classical
mixed state, than this ``something'' gives us information about the
transition probability.  If not, then we cannot distinguish the
superposition from a mixed state.

\subsection{The problem:
generalization to relativistic quasi-classical gravity}

Therefore, this transition probability is very important.  If it is
not observable, what distinguishes the theory from a classical
statistical theory?  Is a theory which does not allow to distinguish a
superposition from a classical mixed state worth to be named ``quantum
theory''?  This seems questionable.  Thus, the assumption that this
transition probability remains observable in full, relativistic
quantum gravity seems to be a very natural one.

Now, to compute the transition probability, we need the scalar product
$\langle\varphi^1|\varphi^2\rangle$.  For the derivation of this
formula, we have used only very few fundamental principles.
Therefore, it seems reasonable to assume that this formula may be
generalized.  We make the following hypothesis: {\em The scalar
product $\langle\varphi^1|\varphi^2\rangle$ is well-defined in
relativistic quantum gravity.\/}

Now, let's consider how to generalize it into the relativistic domain.
The basic states of the source particle
$\psi^1(x)\approx\delta(x-x_1)$ resp. $\psi^1(x)\approx\delta(x-x_1)$
we generalize into relativistic gravitational fields $g^1_{ij}(x)$
resp. $g^2_{ij}(x)$.  For $\varphi(x)$ we have to solve now, instead
of the classical Schr\"odinger equation, a similar wave equation on
these background metrics.  We ignore the related field-theoretical
problems with particle creation and so on and assume that the field
equations may be solved without problems.  Thus, we obtain two
solutions $\varphi^1(x)$ resp. $\varphi^2(x)$ for the two
gravitational fields.

Now let's consider the computation of the scalar product
$\langle\varphi^1|\varphi^2\rangle$, using the following naive formula
as a base:

\[\langle\varphi^1|\varphi^2\rangle=\int\bar{\varphi}^1(x)\varphi^2(x)dx^3\]

The point is that this integral simply cannot be defined from point of
view of classical general relativity.  Indeed, if $g^1_{ij}(x)$ is a
solution of the Einstein equations, we may apply an arbitrary
transformation of coordinates and obtain the same solution in other
coordinates.  Now, if we apply such a transformation to $g^1_{ij}(x)$,
the same transformation has to be applied to $\varphi^1(x)$ too to
obtain the same solution in the other coordinates.  But nothing
requires to apply the same transformation to $g^2_{ij}(x)$ and
$\varphi^2(x)$.  Now, if we apply a coordinate transformation to
$\varphi^1(x)$, but not to $\varphi^2(x)$, the result of the integral
changes in a completely arbitrary way.

The same result may be formulated in another way: the functions
$\varphi^1(x)$ resp. $\varphi^2(x)$ are defined on different manifolds
-- the manifolds defined by the spacetime metrics $g^1_{ij}(x)$
resp. $g^2_{ij}(x)$.  And scalar products between functions on
different manifolds are simply undefined.

A third way to formulate this result is that this scalar product is
only c-covariant but not q-covariant.  Indeed, the integral does not
change if we apply the same diffeomorphism to above configurations,
but changes if we apply different diffeomorphisms to above
configurations.  Therefore, it cannot be observable in q-covariant
quantum GR.

A consequence of this situation is that GR is completely unable to
make predictions for the scalar product, even if the scalar product is
given for some initial values.  Indeed, assume we have fixed some
initial values $g^1_{ij}(x)$, $\varphi^1(x)$, $g^2_{ij}(x)$,
$\varphi^2(x)|_{x^0=0}$ and an appropriate number of derivatives.
Thus, for the initial values we have restricted the freedom of choice
of diffeomorphisms.  But this does not help: there are diffeomorphisms
which are identical for the initial values, with all derivatives.  And
the Einstein equations define the solution only modulo arbitrary
diffeomorphisms.

What seems to be even more serious is that the problem appears already
in the non-relativistic limit.  In Schr\"odinger theory, the scalar
product is well-defined.  But the gravitational field may be as close
as possible to the non-relativistic situation, the scalar product
remains completely undefined in relativistic theory.  It seems
therefore highly problematic to obtain Schr\"odinger theory as the
non-relativistic limit of a q-covariant theory of quantum gravity.

\subsection{The solution: a fixed space-time background}

Let's look now how this problem is solved in GET.  We have some
well-defined Newtonian background which is common for all field
configurations.  The additional term in the Lagrangian does not only
give some additional term in the Einstein equations, but breaks
relativistic diffeomorphism invariance.  We have four additional
equations -- the harmonic coordinate equations.  They define the
solution uniquely, not only modulo diffeomorphism.

Of course, ``uniquely'' is also a relative notion, it means relative
to the Newtonian background.  We can use the covariant formulation of
GET, with a covariant equation for the variables used to define the
background.  The conceptual difference is that this background is
common for all solutions.  The resulting quantum theory is not
q-covariant, but only c-covariant.  The relative position between a
general field configuration and the common Newtonian background
defines relative positions between different field configurations.
This allows to define scalar products as well as the notion ``the same
diffeomorphism'' which is necessary to define c-covariance.

Are there other ways to define the evolution of these scalar products?
No.  By accepting the existence and observability of the scalar
product we have de facto accepted a common background manifold with
preferred coordinates for all semi-classical gravitational fields.
Indeed, we need only a few number of simple and natural restrictions
to obtain a common position measurement for all gravitational fields.

Assume as before that we have two configurations of gravitational
fields $g^1_{ij}(x)$, $g^2_{ij}(x)$ on manifolds $M_1$ resp.  $M_2$
and some Hilbert space of appropriate wave functions $H(M_1)$ and
$H(M_2)$.  Now, we assume the element $\varphi^1(x)\in H(M_1)$ has
well-defined ``scalar products'' with all elements of $H(M_2)$.  But
in this case it defines a linear functional on $H(M_2)$.  A linear
functional on $H(M_2)$ uniquely defines an element of $H(M_2)$.  This
construction defines a map $H(M_1)\to H(M_2)$.

Now, this map seems to be the appropriate place to define additional
natural requirements.  First, we need transitivity.  If there are
three spaces, the map $H(M_1)\to H(M_3)$ should be the same as the
composition $H(M_1)\to H(M_2)\to H(M_3)$.  As a special case $M_1 =
M_3$ we obtain that the map $H(M_2)\to H(M_1)$ is the inverse of
$H(M_1)\to H(M_2)$.  Another property is that they are
norm-preserving.  This is required to have a consistent probability
interpretation.  Above restrictions may be justified in the same way
we have justified the existence of the scalar product itself: these
properties are fulfilled in Schr\"odinger theory.

But, once we have a norm-preserving map $H(M_1)\to H(M_2)$, we can use
it to transfer a measurement from $M_2$ to $M_1$.  Especially, we can
transfer the position measurement for the manifold $M_2$ to $H(M_1)$.
Now, we can choose an arbitrary solution as a reference solution and
transfer its position measurement to all other states as the common
background.  Thus, we obtain a common background manifold for all
field configurations.  We are de-facto back to the scheme we use in
GET, with a fixed common space-time background.

Thus, if we follow the relativistic paradigm and develop a q-covariant
quantum theory of gravity, important observables of non-relativistic
quantum gravity remain undefined.  If we assume that they are
well-defined, we have to reject the relativistic paradigm and to
introduce a common background manifold into the theory.  This
consideration justifies the introduction of a fixed space-time
background into GET.

\subsection{Comparison of Regge calculus and dynamical triangulation}

It is interesting to compare two well-known discrete approaches to
quantum gravity -- the Regge calculus \cite{Regge} and dynamical
triangulations (DT) from point of view of scalar product.  In the
Regge calculus, we have a fixed grid and the geometry is described by
the edge lengths of the grid.  In contrast, in DT the edge lengths are
fixed but the triangulation varies.

Now, if we have some discrete functions defined for different
geometries in the Regge calculus, we can define their scalar product
without problem -- we have the same grid as the base, therefore, for
each point of one geometry we have a well-defined notion of ``the same
point'' on the other geometry.  Therefore, the Regge calculus is a
nice example of a discrete c-covariant theory.  Instead, in DT we do
not have such a possibility.  The scalar product between discrete
functions on different triangulations is meaningless.  Therefore, it
is an example of a discrete q-covariant approach.

Thus, according to the ideology presented here, DT should be the
``correct'' way to quantize geometry in a diffeomorphism-invariant
way, but leads to problems with the classical limit, while it should
be reverse for the Regge calculus.  Indeed, the review
\cite{Catterall} shows exactly these properties of the two approaches:
the Regge calculus with fixed grid contains steps of freedom which are
unphysical from point of view of relativity, especially modes
corresponding to general coordinate transformations, but it
``possesses a weak field expansion in which contact can be made with
continuum perturbation theory''.

It is also noted that, at least for 2D, ``the DT method affords a good
prescription for regulating quantum gravity''.  But ``there is no weak
coupling limit in which contact can be made with continuum perturbation
theory.  Indeed, the attractive feature of this formulation -- that it
is purely geometric, making no reference to coordinates and metric
tensors also poses a problem; how do such classical quantities emerge
from the model at large distance''.

I hope, these remarks help to clarify my understanding of the role of
the scalar product problem: it does not claim that q-covariant
theories are impossible.  Instead, DT provides an example of the
regularized version of such a theory.  It also does not claim that by
accepting the existence of a scalar product we immediately end with
ether theory: there is a certain difference between Regge QG and our
ether approach, especially in our ether approach the grid is not
fixed, but moves, and the position of the grid nodes are steps of
freedom of the ether approach.

The point is that a q-covariant approach has to be rejected because of
the failure to define a scalar product between functions defined on
different solutions, because such scalar products are necessary in the
non-relativistic limit.

\section{Realism as a methodological concept}\label{EPR}

\newtheorem{epr-axiom}{Axiom}

In \S{ } \ref{Bell} we have not considered the EPR criterion of
reality.  There was no necessity for this, because we have considered
it as part of common sense.  To separate one of the common sense
principles used in this proof and to name it ``EPR criterion'' is not
necessary in our approach.  Instead, it is part of the destruction
strategy we have considered in section \S{ } \ref{destruction}.

Unfortunately, this destruction strategy was already successful.
Therefore, it seems necessary to consider the part of common sense
which has been named ``EPR criterion of reality'' and questioned in
more detail.

\subsection{Principles of different importance}

If we have a contradiction between theory and experiment, there are
always different parts of the theory which may be blamed for the
problem.  But often it is not too difficult to find the critical part.
Usually it is very helpful that different parts are not on the same
level of fundamentality.  We usually can distinguish more and less
fundamental parts of the theory, and in case of conflict we usually
blame the less fundamental parts to be the cause of the problem.
Let's consider, for example, the dark matter problem.  We observe a
difference between the Einstein equations and observation:

\[ G^{\mu\nu} \neq 8\pi G T^{\mu\nu}_{obs} \]

In this case, we do not reject GR because it's main equation is
falsified by observation.  Instead, we simply define the
energy-momentum tensor of ``dark matter'' as

\[ T^{\mu\nu}_{dark} = G^{\mu\nu} - 8\pi G T^{\mu\nu}_{obs} \]

and obtain that the Einstein equations are fulfilled.  In this case,
our existing theory of matter is considered to be less fundamental.
In this case, this seems to be a reasonable choice.

We argue here that the situation is different for the violation of
Bell's inequality.  The other principles involved in the proof of
Bell's inequality, which we denote here as the principles of realism
and which include the EPR criterion of reality, are more fundamental
than the particular assumption about space-time symmetry known as
relativity.  Especially, we argue that these other principles are
fundamental methodological principles, part of the methodological
foundations of science.  Thus, they are important as in physics, as in
other sciences.

Note that this argument is only additional support for our
argumentation.  There is already another argumentation which is
completely sufficient for a unique decision in favor of realism:
There is independent evidence against relativity -- their problems
with quantum gravity, especially the problem of time (appendix
\ref{time}).  Moreover, there is a viable competitor of relativity --
ether theory -- which has been developed to solve these other
problems.  Instead, nobody has proposed a theory which, for
independent reasons, rejects realism.  Moreover, there is simply
no independent evidence against realism: as we have seen, quantum
theory is compatible with realism.

\subsection{A definition of reality and causal influence}

Let's try to define classical realism in a way which allows a strong
mathematical proof of Bell's inequality based on this notion of
realism.  Realism in the common sense proposes the existence of an
observer-external reality which {\em exists\/} independent of our
observation.  The results of observations may be results of complex
interactions between reality and observer, nothing requires a
possibility of direct observation.  To define realism we need at least
the following three entities: of course the {\bf observables}, but
also the {\bf decisions} of the experimenters what to measure, and
last not least the {\bf reality}.  But that's all we need:

\begin{epr-axiom}[reality]

Assume we have an experiment described by observables $X$ with
the observable probability distribution $\rho_X(X,x)dX$.  It depends
on a set of control parameters $x$ which describe the
experimental setup (the decisions of experimenters).

A theory is realistic if it describes such probability distributions
based on a notion of {\bf reality} -- a space $\Lambda$ (reality) with
probability distribution $\rho_\lambda(\lambda)d\lambda$ -- and a {\bf
realistic explanation} -- a function $X(x,\lambda)$ -- so that for a
test function f

\[ \int f(X)\rho_X(X,x)dX=\int f(X(x,\lambda))\rho(\lambda) d\lambda \]

\end{epr-axiom}

This formal definition is in quite good agreement with the common
sense idea that reality $\lambda$ exists independent of our decisions
$x$: the probability distribution $\rho(\lambda)$ indeed does not
depend on $x$.  But it already incorporates the insight that there is
no pure observation, that our observations are only the result of
complex interactions between observer and reality.  An argumentation
that classical realism is invalid because observations are only the
result of such interactions is, therefore, invalid: this possibility
is already part of classical realism.

Note that already on this level, without any relation to space-time,
we can define causal influences in a natural way:

\begin{epr-axiom}[causal influence]
If in a realistic theory an observable X depends on a control
parameter x in the realistic explanation $X(x,\lambda)$, then
we have a {\bf causal influence} of x on X.
\end{epr-axiom}

\subsection{Bell's inequality as a fundamental property}

These definitions are already sufficient for the proof of Bell's
inequality.  In the case of Bell's inequality, the control parameters
are the questions: your question $a$ to Alice and your friends
question b to Bob.  The observables are their answers A and B.  Using
the definition of realism we obtain the existence of two functions:

\[ A=A(a,b,\lambda)=\pm 1; \hspace{2cm}  B=B(a,b,\lambda)=\pm 1  \]

We also obtain the expectation value for the product AB as

\[P(a,b)=\int\rho(\lambda)A(a,b,\lambda)B(a,b,\lambda)d\lambda.\]

Now we have to consider causality.  If there is no causal influence of
the decision b on the result A, then we have $A=A(a,\lambda)$ and
resp. $B=B(b,\lambda)$.  Thus, we obtain

\[P(a,b)=\int\rho(\lambda)A(a,\lambda)B(b,\lambda)d\lambda\]

which is simply formula (2) of \cite{BellTheorem}.  After this, Bell's
inequality follows as derived in \cite{BellTheorem}.

As a consequence, we obtain the proof of Bell's inequality on a level
where even the {\em existence\/} of something like space-time has not
been mentioned, and therefore in a space-time independent form: if it
is violated, this proves the existence of causal influences $(a\to B)$
or $(b\to A)$.

\subsection{The methodological character of this definition}

A remarkable property of this definition is its unfalsifiable
character.  Whatever we observe, it is possible to describe it using a
probability distribution $\rho_X(X,x)dX$.  Whatever this probability
distribution is, we can always construct a realistic theory which
leads to this distribution -- all we have to do is to use a
(sufficiently artificial) functional space to describe the reality.
Especially, reality may be described simply by the measure
$\rho_X(X,x)dX$ itself.\footnote{As the many worlds interpretation, as
Bohmian mechanics may be considered as realistic theories obtained as
variants of this ``cheap'' way -- they simply accept the wave function
as reality.}

The reasonable question is about the purpose of this definition if it
is unfalsifiable.  Now, there is a surprisingly simple answer: realism
is simply a {\em methodological rule\/}.  It {\em enforces\/} to
describe certain parts of the theory as {\em really existing\/}.  As
well, the subsequent definition of a causal influence is also
unfalsifiable.  The purpose of this definition is, as well, to {\em
enforce\/} to name some relations {\em causal influence\/}.  In other
words, this definition of realism and causality enforces ontological
clarity.  A realistic theory is a theory where we are forced to name
some things real and some influences causal, even if this violates our
metaphysical prejudices or principles of the theory we prefer.

Especially this {\em definition\/} of realism and causality, a
definition already enforces that any violation of Bell's theorem
should be explained by causal influences -- or $a\to B$ or $b\to A$.

\subsection{Causality requires a preferred frame}

It does not follow from the {\em definition\/} that these causal
influences happen in a preferred frame.  To prove the existence of a
preferred frame we need a little bit more.  First, the connection
between causality and space-time.  Until now, even the existence of
something like a space-time has not been mentioned.  Only now we have
to define causality on space-time as a relation $x\to y$ between
space-time events x, y: $x\to y$ if there exists some $a\to A$ so that
the decision a is localized at x and the observation A localized at y.
Moreover, we need the most important property of causality: the causal
order along a world line and the absence of causal loops.

\begin{epr-axiom}[space-time causality]
Causality defines a partial order $x\to y$ on space-time with the
property that on time-like trajectories $\gamma(t)$ we have
$\gamma(t_0)\to\gamma(t_1)$ if $t_0<t_1$.
\end{epr-axiom}

Now, we can simply prove the existence of a preferred foliation:
\footnote{That the existence of a preferred foliation follow is
nothing new. Valentini \cite{Valentini92} suggests ``that a preferred
foliation of spacetime could arise from the existence of nonlocal
hidden-variables'' \cite{Isham}.  Bell himself concludes \cite{Bell1}:
``the cheapest resolution is something like going back to relativity
as it was before Einstein, when people like Lorentz and Poincare
thought that there was an aether --- a preferred frame of reference
--- but that our measuring instruments were distorted by motion in
such a way that we could no detect motion through the aether.''}

\begin{theorem}[existence of a preferred foliation]
If for all pairs of events Bell's inequality may be violated, then
there exists a preferred foliation.  It is defined by the property
that if $x\to y$ than $T(x)<T(y)$ for the function T(x) which defines
this foliation.
\end{theorem}

Proof: Let's define the foliation as a time-like function $T(x^i,t)$.
For this purpose, we set $T(0,t_0)=t_0$ and define the points
contemporary to $A=(0,0,0,t_0)$ on the line $B(t)=(x^1,x^2,x^3,t)$.
We use the Dirichlet algorithm on the line $B(t)$.  We start with a
large enough interval $(a_0,b_0)$ so that there exists causal
influences $B(a_0)\to A \to B(b_0)$.  Assume at step n we have found
an interval $B(a_n)\to A \to B(b_n)$ with $|b_n-a_n|<2^{-n}|b_0-a_0|$.
Now, we consider the element $B(h = (b_n+a_n)/2)$.  Then we observe a
violation of Bell's inequality between A and $B(h)$.  It follows from
Bell's theorem that there should be $A\to B(h)$ or $B(h)\to A$.  In
the first case, we set $a_{n+1}=a_n,b_{n+1}=h$, else
$a_{n+1}=h,b_{n+1}=b_n$.  In above cases we have found an interval
with $B(a_{n+1})\to A \to B(b_{n+1})$ with
$|b_{n+1}-a_{n+1}|<2^{-n-1}|b_0-a_0|$.  Therefore, we have a limit
$l=\lim a_n=\lim b_n$.  This limit defines a function
$T(x^1,x^2,x^3,t_0)=l$.

To prove that the function $T(x)$ is correctly defined, Lipschitz
continuous, and that the definition of the foliation does not depend
on the choice of coordinates is straightforward.  What we need is that
in every environment of $B(l)$ we have points $a_n$ with $B(a_n)\to A$
as well as points $b_n$ with $A\to B(b_n)$, the non-existence of
causal loops, and the existence of causal ordering $B(a)\to
B(b)\Leftrightarrow a<b$ on time-like trajectories $B(t)$.

\subsection{Relation between our definition and the EPR criterion}

Let's consider now the difference between this definition of realism
and the EPR criterion of reality \cite{EPR}:

\begin{quote}
If, without in any way disturbing a system, we can predict with
certainty ... the value of a physical quantity, than there exists an
element of physical reality corresponding to this physical quantity.
\end{quote}

Now, this is a natural consequence of our definition of realism: We
have no ``disturbance'', thus, no dependence of A on b:
$A=A(a,\lambda)$.  If it is a prediction, because we have no influence
backward in time, we have no dependence of B on a: $B=B(b,\lambda)$.
It is a prediction with certainty, thus, these functions are identical
as functions: $A(.,\lambda) = B(.,\lambda)$.  Last not least,
$\lambda$ is the ``element of reality'' and the function
$A(.,\lambda)$ describes the correspondence to the physical quantity
A.

The advantages of our new definition are, in our opinion, the
following: we have a general definition, while the EPR mentions a
special situation -- a correlation which allows to predict something
with certainty, and we do not depend on the notion of space-time,
while the EPR criterion includes an implicit reference to time
(``predict'').  Moreover, the formal character of the definition
allows to show its methodological character: it does not restrict
physical theories, but restricts our way to talk about them.

\subsection{Methodological principles as the most fundamental part of
science}

Once we defend realism as a fundamental methodological principle it
seems useful to look how other fundamental principles of science may
be defended.  There is another such fundamental methodological
principle -- classical logic, especially the law that there should be
no contradictions, nor in the theory, nor between theory and
observation.  As our definition of realism and causality these
principles are unfalsifiable themself -- simply because the principle
of falsification itself relies on classical logic.\footnote{If they
are false, then the method of falsification is false too, therefore,
cannot be used.  Even if the use of rational arguments, especially the
``therefore'' in the last sentence, is also unjustified, this argument
seems to show that an experimental falsification of classical logic is
impossible.}  Therefore, other arguments have to be used to defend
them.  In this context, it is interesting how Popper defends classical
logic against ``dialectical logic'' (\cite{PopperCR}, p.316):

\begin{quote}
Dialecticians say that contradictions are fruitful, or fertile, or
productive of progress, and we have admitted that this is, in a sense,
true.  It is true, however, only so long as we are determined not to
put up with contradictions, and to change any theory which involves
contradictions; in other words never to accept a contradiction: it is
solely due to this determination of ours that criticism, i.e. the
pointing out of contradictions, induces us to change our theories, and
thereby to progress.  It cannot be emphasized too strongly that if we
change this attitude, and decide to put up with contradictions, then
contradictions must at once lose any kind of fertility.
\end{quote}

Thus, the point of the argumentation is not to prove that there can be
no contradictions.  The basic idea is that we have to consider not the
hypothesis itself, but their influence on the future development of
science.  That means, Popper defends classical logic as a
methodological rule of science.

The main advantage of this argumentation is that we do not have to
rely on ``common sense'' -- a notion which has a bad name in current
science and is usually compared with flat Earth theory.  Classical
logic is not a particular common sense theory like flat Earth theory
which may be false, but defines the scientific method, therefore, if
we reject classical logic, we simply reject the scientific method.

Our notion of realism is fertile in the same sense as classical logic.
It is the rule to search for realistic, causal explanations for
observable correlations.  A non-trivial, unexplained correlation plays
the same role as the contradiction in logic: it defines a scientific
problem.  We have to include a realistic explanation into our theory.
Nobody forces us to search for explanations, it is only our own {\em
methodological decision} not to accept unexplained correlations, and
to accept only a realistic explanation.  If we give up the search for
realistic explanations, we loose an important way to reach scientific
progress.

\subsection{The methodological role of Lorentz symmetry}

The great importance of Lorentz symmetry in modern physics is often
presented as if it is a decisive argument against a preferred frame.
But this suggests that Lorentz symmetry would have been less important
in the Lorentz ether.  Is there any evidence for this claim?  I have
never seen any justification for this assumption.  It is simply
claimed, without justification, that people would have been less eager
to search for relativistic symmetry.  The reverse may be closer to
truth.  Instead, with the Lorentz ether as the leading ideology,
people would have tried to detect hidden Lorentz symmetry in usual
condensed matter theory.

Moreover, without doubt any part of the hidden variables which can be
made Lorentz-covariant would have been made Lorentz-covariant.  An
example are the equations of GET presented here.  The new equation for
the preferred coordinates is a nice, well-known relativistic equation
-- the harmonic equation.  Moreover, the thesis is in obvious
contradiction to the history of the Lorentz ether.  In the context of
the Lorentz ether, by Poincare, the program to make all physical
theories Lorentz-invariant has been proposed in general and realized
for kinematics.  The decision to reject the existence of a preferred
frame made by Einstein was in no way necessary for the development of
this program.

The point is that to require a particular symmetry is not a general
methodological rule of scientific research.  At best there is the
related methodological rule to search for symmetries in general.  But
even this rule seems much less fundamental than classical logic and
realism: last not least, we search for symmetries in reality.
Symmetries are a powerful {\em tool\/} to study realistic theories, to
detect contradictions in such theories or between theory and
experiment -- but only a tool, in no way a fundamental principle.

\subsection{Discussion}

As presented here, the preferred frame is the unavoidable consequence
of the violation of Bell's inequality.  Relativity is falsified by
Aspect's experiment, and its current status should be rejected as an
immunization.  This is so obvious that it becomes problematic to
explain the unreasonable decision of mainstream science to reject
realism.  But there are several factors which may be blamed here:

\begin{itemize}

\item The absence of a reasonable theory of gravity with preferred
frame.  This problem is solved now by GET.

\item The widely accepted belief, based on von Neumann's \cite{Neumann} 
theorem, that hidden variable theories are impossible.  This was
justified at the time the EPR criterion was proposed, but many seem to
believe it even today.

\item The ignorance of Bohmian mechanics {\em because\/} it requires a
preferred frame.

\item The extreme positivism and subjectivism during the foundational
period of quantum theory.

\item The general ignorance of fundamental problems of quantum
theory today.

\end{itemize}

But the most important explanation seems to be Kuhn's theory of
paradigm shifts \cite{Kuhn}.  According to Kuhn, paradigms are never
falsified by experiments.  A paradigm may be rejected only if a new
paradigm appears.  Until now, no alternative paradigm has been
proposed, therefore, to preserve the relativistic paradigm was
justified -- in full agreement with Kuhn's paradigm shifts.

We propose here a new paradigm -- a return to classical Newtonian
space-time and ether theory.  With this paradigm as a competitor of
the relativistic paradigm it is no longer necessary to reject realism
or causality.

\section{Bohmian mechanics} \label{Bohm}

An essential property of non-relativistic Schr\"odinger theory is the
existence of a simple deterministic interpretation -- Bohmian
mechanics (BM).  We refer to this theory in our proof that EPR realism
is not in conflict with non-relativistic quantum theory.
Unfortunately, BM is widely ignored.  The main reason for this
ignorance seems to be that it requires a preferred frame -- thus, a
feature which makes it particularly attractive in the context of GET.
Therefore, it seems reasonable to consider the basic features of BM
here.

\subsection{Simplicity of Bohmian mechanics}

BM may be considered as a straightforward way to complete quantum
mechanics.  In BM, we have two entities: the ``guiding wave''
$\Psi(q)$ defined on the configuration space which fulfills the
classical Schr\"odinger equation

\[ i \partial_t \Psi = H \Psi \]

and the configuration $Q(t)$ which fulfills the so-called ``guiding
equation''.  This guiding equation may be obtained in a
straightforward way from quantum mechanics.  The basic observation is
the following: quantum mechanics provides us with a probability
current $j^i(q)$ as well as with a probability density
$\rho(q)=\Psi^*(q) \Psi(q)$.  In classical mechanics they are related
by $j^i(q) = \rho(q) v^i(q)$.  Now it requires no great
imagination to write the guiding equation

\[ {d Q\over dt} = v^i = {j^i \over \rho} \]

This defines the evolution of the state.  Now, if in initially the
state is in the so-called ``quantum equilibrium'' $\rho(q)$, then it
remains in this state.  This follows from the continuity equation

\[ \partial_t \rho(q,t) + \partial_i j^i(q,t) = 0 \]

That's already all what is necessary.  There is no need for further
axioms.  Therefore, all we need for the definition of Bohmian
mechanics is the quantum probability current.  For example, in
non-relativistic multi-particle theory

\[ H = - \sum_{k=1}^N {\hbar^2\over m_k^2} \nabla_k^2 + V(q_1,\ldots,q_N) \]

this probability current is given by

\[ j_k = {\hbar\over m_k} \Im (\psi^*\nabla_k\psi) \]

Therefore, we obtain the guiding equation

\[ {d Q_k\over dt} = {\hbar\over m_k} \Im {\nabla_k\psi\over\psi} \]

\subsection{Clarity of the interpretation}

The first thing we have to note here is the ontological clarity.  To
quote Bell (\cite{Bell}, p.191):

\begin{quote}
Is it not clear from the smallness of the scintillation on the screen
that we have to do with a particle?  And is it not clear, from the
diffraction and interference patterns, that the motion of the particle
is directed by a wave? ... This idea seems to me so natural and
simple, to resolve the wave-particle dilemma in such a clear and
ordinary way, that it is a great mystery to me that it was so
generally ignored. ...
\end{quote}

This solution of the wave-particle confusion not the main point: in BM
there is also nothing strange with Schr\"odinger's cat.  The wave
function of the cat remains in its superpositional state, but the
actual cat is in a well-defined state.  ``There is no need in this
picture to divide the world into `quantum' and `classical' parts. For
the necessary `classical terms' are available already for individual
particles (their actual positions) and so also for macroscopic
assemblies of particles.'' (\cite{Bell}, p.192)

Another interesting question is worth to be mentioned: why are the
states we observe in quantum equilibrium?  This question has an
interesting answer: decoherence.  ``One of the best descriptions of
decoherence, though not he word itself, can be found in Bohm's 1952
`hidden variables' paper \cite{Bohm}. We wish to emphasize, however,
that while decoherence plays a crucial role in the very formulation of
the various interpretations of quantum theory loosely called
decoherence theories, its role in Bohmian mechanics is of quite
different character: For Bohmian mechanics decoherence is purely
phenomenological -- it plays no role whatsoever in the formulation (or
interpretation) of the theory itself'' \cite{Duerr}.

The most important property of BM is it's compatibility with classical
principles: the EPR criterion of reality, classical causality,
determinism.  Let's quote again Bell (\cite{Bell}, p.163):

\begin{quote}
It is easy to find good reasons for disliking the de Broglie-Bohm
picture. Neither de Broglie nor Bohm liked it very much; for both of
them it was only a point of departure.  Einstein also did not like it
very much. He found it `too cheap', although, as Born remarked, `it
was quite in line with his own ideas'.  But like it or lump it, it is
perfectly conclusive as a counter example to the idea that vagueness,
subjectivity, or indeterminism, are forced on us by the experimental
facts covered by non-relativistic quantum mechanics.
\end{quote}

\subsection{Relativistic generalization}

Let's consider now the main point why many researchers dislike BM --
its relativistic generalization.  The same basic scheme works as well
in relativistic theory and field theory.  For example, for multiple
Dirac particles Bohm \cite{BohmHiley} has proposed the following
guiding equation:

\[ {\bf v}_k = { \psi^{+} {\bf \alpha}_k \psi\over  \psi^+ \psi} \]

For the general case of quantum field theory, we have to accept the
lectures of quantum field theory what is the appropriate notion of the
wave function: ``Certainly the Maxwell field is not the wave function
of the photon, and for reasons that Dirac himself pointed out, the
Klein-Gordon fields we use for pions and Higgs bosons could not be the
wave functions of the bosons.  In its mature form, the idea of quantum
field theory is that quantum fields are the basis ingredients of the
universe, and particles are just bundles of energy and momentum of the
fields.  In a relativistic theory the wave function is a functional of
these fields, not a function of particle coordinates''
\cite{Weinberg}.  Thus, it does not make sense to search for a guiding
equations for particles in the general case, and we have to consider
Bohmian field theory \cite{BohmHiley} where we obtain a guiding
equation for generalized coordinates -- the field configuration.

Thus, the generalization itself is not problematic.  It is an
essential property of this generalization -- that it has an explicit
preferred frame on the fundamental level.  The predictions are
nonetheless Lorentz-invariant.  For example, $\psi^+ \psi$ is an
equivariant ensemble density {\em in the chosen reference frame.\/} It
reproduces the quantum predictions in this frame.  These predictions
don't contain a trace of the preferred frame.  Lorentz invariance
holds on the observational, but not on the fundamental level.  The
4-tuple $(\psi^+ \psi, \psi^+ {\bf \alpha}_k \psi)$ is not a 4-vector
for $N>1$.

This is not an accident. ``There does not in general exist a
probability measure P on N-paths for which the distribution of
crossing $\rho^\Sigma$ agrees with the quantum mechanical distribution
on all space-like hyper-planes $\Sigma$'' \cite{Berndl}. This assertion
is a more or less immediate consequence of Bell's inequality: by means
of a suitable placement of appropriate Stern-Gerlach magnets the
inconsistent joint spin correlations can be transformed to (the same)
inconsistent spatial correlations for particles at different times
\cite{Berndl}.  Thus, we have the probability measure $\rho=|\psi|^2$
only in one frame.  But this measure in just {\em one\/} frame is
sufficient to derive the quantum mechanical predictions for
observations at different times.

\subsection{Discussion}

The fact that Lorentz invariance does not hold on the fundamental
level is often considered as a decisive argument against BM.  But from
point of view of ether theory this becomes a virtue rather than a
vice: every argument in favour of BM becomes an argument in favour of
the preferred frame we use in ether theory.  To use the argument
``there is no fundamental Lorentz-invariance'' against BM in this
context would be simply circular reasoning -- a main concept of ether
theory is as well that there is no fundamental Lorentz-invariance.

Thus, BM gives additional support for one of the main ingredients of
GET -- the preferred absolute time.  One the other hand, GET gives
support to BM -- it shows a way to generalize BM to gravity.  We do
not have to try to find Lorentz-invariant versions of BM, as tried,
for example, in \cite{Berndl}.  Instead, we can apply BM as it is,
with a preferred frame, in GET or, even better, in an atomic ether
theory.

\end{appendix}

\end{document}